\documentclass[preprint,12pt]{elsarticle}
\usepackage{caption}
\usepackage{array}
\newcolumntype{L}[1]{>{\raggedright\arraybackslash}p{#1}}

\usepackage{algpseudocode}
\usepackage{amsmath, amssymb, amsfonts, mathtools}
\usepackage{enumitem}

\usepackage[ruled,vlined]{algorithm2e}   % replaces algorithm + algpseudocode combo cleanly

\usepackage{graphicx}
\usepackage{placeins}        % for \FloatBarrier
\usepackage{textcomp}        % for special symbols like \textdegree

\usepackage{booktabs}
\usepackage{multirow}
\usepackage{tabularx}
\usepackage{array}
\newcolumntype{Y}{>{\centering\arraybackslash}X}  % for centered X columns in tabularx

\usepackage{enumitem}        % for compact and custom enumerations
\usepackage{enumerate}       % traditional enumerations

\usepackage{hyperref}
\hypersetup{
    colorlinks=true,
    linkcolor=blue,
    citecolor=blue,
    urlcolor=blue
}

\usepackage{amsthm}

\usepackage{footnote}

\journal{Journal of Energy Storage}

\begin{document}

\begin{frontmatter}

\title{Health feature extraction from 
battery energy storage system field fault data}

\author[1]{Clement Wong}
\author[1]{Andrew Weng}
\author[1]{Xin Hui Ooi}
\author[1]{Zhiwen Wan}
\author[2]{Jeesoon Choi}
\author[2]{Seung Yoon Yang}
\author[2]{Heejun Jin}
\author[1]{Jason Siegel}
\author[1]{Anna Stefanopoulou}

\affiliation[1]{organization={Department of Mechanical Engineering, University of Michigan},
               addressline={1231 Beal Ave},
               city={Ann Arbor},
               state={Michigan},
               postcode={48109},
               country={USA}}

\affiliation[2]{organization={LG Energy Solution},
               % addressline={},
               city={Seoul},
               % postcode={34122},
               country={South Korea}}

%% Abstract
\begin{abstract}
Health monitoring methods are critical for lithium-ion battery modules connected to the grid to prevent faults that can lead to catastrophic events. However, assessing the health of cells in modules from their operational data presents challenges including variable operating conditions, which directly confound health features, and sparse sensing in the modules, particularly within cells in parallel, which prevents observing critical states of individual cells. Here, we present a framework for extracting and calibrating health features for battery modules from their operational data to identify discriminative features for separating faulty parallel-connected cell groups within the modules. We applied this framework to operational data from 25 commercial grid-connected lithium-ion Battery Energy Storage System (BESS) modules. Each module consisted of 14 series-connected parallel groups, one of which was confirmed as faulty via post-mortem investigation; in total, the dataset included 25 faulty and 325 non-faulty cell groups. A statistical evaluation of these calibrated features demonstrated that group-level capacity, capacity degradation rate, and dV/dQ peak heights separate faulty parallel-connected cell groups within the modules with statistical significance ($p<0.05$). Conversely, group internal resistance did not ($p>0.05$), indicating that increased resistance was not a primary characteristic of the faults in this dataset. These findings challenge the exclusive reliance on resistance features for fault detection. The observed feature signatures suggest potential failure mechanisms, furthering the understanding of fault behavior in lithium-ion battery modules during field operation. More importantly, this work demonstrates a framework for robustly monitoring the health of cells in lithium-ion battery modules under real-world operations.
\end{abstract}

% %%Graphical abstract
% \begin{graphicalabstract}
% \includegraphics[width=\textwidth, keepaspectratio]{graphical_abstract.pdf}
% \end{graphicalabstract}

% %Research highlights
% \begin{highlights}

% \item Health feature extraction applied to lithium-ion BESS module field data
% \item Gaussian process regression for feature calibration and uncertainty quantification
% \item Comparing features for separating faulty parallel-connected cell groups in modules
% \item Results show degradation of field module and suggest potential failure mechanisms

% \end{highlights}

%% Keywords

\end{frontmatter}

%% Add \usepackage{lineno} before \begin{document} and uncomment 
%% following line to enable line numbers
%% \linenumbers

%% main text
%%

\section{Introduction}

Lithium-ion battery modules are central to many large-scale battery energy storage systems (BESS) that support our grid infrastructure. Battery faults in these modules pose risks, including major property damage and economic losses~\cite{Im2023}. Therefore, to ensure the reliability and safety of these systems, it is critical to detect faults within the battery modules during their field operations to enable preventative actions.

Extensive research has been conducted on fault detection for lithium-ion batteries; however, much of this work has been based in the laboratory setting, and translating these findings to field applications presents many challenges. One significant challenge is that in the field, the fault mechanism is generally unknown \cite{Luder2025}. In laboratory settings, researchers induce abuse conditions, such as mechanical, electrical, or thermal stresses, to study fault behaviors and mechanisms, thereby understanding both the cause and the likely effects on battery behavior~\cite{Chen2019a, Han2020,Yi2013}. In contrast, information regarding the faults in the field and their underlying mechanisms is often unknown, complicating the task of determining how a fault in the field may affect battery behavior and how such effects are reflected in operational data~\cite{Zhao2024, Hu2020}. Another challenge is the variability of operating conditions in real-world systems, which contrasts with the controlled environments of laboratory experiments~\cite{SULZER20211934}. This variability directly affects health features, complicating distinguishing if changes in health features are reflective of changes in degradation versus changes in the operating conditions~\cite{Antti_Field}. Furthermore, unlike high-resolution laboratory instruments, sensors used in the field are typically inexpensive and can introduce significant noise into the data~\cite{SULZER20211934, Antti_Field}. This inherent noise can obscure the subtle, transient electrical and thermal signatures that may indicate the onset of a fault.

The limited number of sensors in a battery system further complicates fault detection. In most commercial battery systems, there are only voltage, current, and temperature sensors, which may not be effective for capturing faults as compared to other sensors such as gas, force, or internal strain sensors~\cite{Cai2021, Fan2025, Koch2018}.  Even if the fault can be directly reflected in voltage, current, and temperature, its signature may be difficult to observe due to the sparse placement of the sensors~\cite {Lin2020-tb}. Typically, voltage and current are measured only at the level of each parallel-connected cell group, and the number of temperature sensors in a module is often fewer than the number of series-connected cells~\cite{Pozzato2023, Schaeffer2024, Figgener2024}. Because there is not a dedicated sensor for each cell, the states of individual cells within the system cannot be directly observed, making it difficult to pinpoint the subtle, cell-level signals that could indicate an emerging fault~\cite{Lin2020-tb}.

With the low probability of fault occurrence, data with faults are very hard to obtain, hindering the development and validation of methods to detect faults in battery modules in the field~\cite{Ward2022, Hu2020}. Recent work by Schaeffer et al. represents a significant step forward by releasing a public dataset of real-world battery systems with faulty behavior, providing a valuable resource for the research community \cite{Schaeffer2024}. However, the scope of their analysis presents clear opportunities for further investigation. First, the systems in their dataset consisted solely of cells connected in series and do not contain cells in parallel connection, and thus do not address the fault dynamics of the many commercial systems that utilize parallel-connected cell groups. Furthermore, their study focuses exclusively on internal resistance as a health indicator, leaving open the question of whether other health features could serve as potentially more effective indicators for fault detection in the field~\cite{Attia2022,Berecibar2016b,Xiong2018}. Therefore, there remains a critical need to analyze faults in systems with parallel cell connections and to evaluate a broader suite of health features for their ability to indicate real-world faults.

In this work, we analyze a unique field dataset from 25 commercial lithium-ion BESS modules; each module consisted of 14 series-connected parallel groups, one of which was identified as faulty via post-mortem analysis. The dataset provides 1 Hz operational data for a period of 1 to 6 months leading up to the fault, generating over 183 million data rows.  From this real-world operational data, we develop and apply algorithms to extract health features—including capacity, internal resistance, and differential voltage analysis (DVA) features— for each parallel-connected cell group (Section~\ref{sec:feature_extraction}). We then 
employ Gaussian Process (GP) regression to calibrate the health features, separating the confounding effects of operating conditions to isolate the time-dependent trend that reflects the battery's true underlying degradation trajectory (Section~\ref{sec:feature_sensitivity_to_operation_condition}). By decoupling operational sensitivities, our method enables continuous health monitoring whenever features are extractable, regardless of the specific operating conditions. Finally, we statistically evaluate the effectiveness of these features in distinguishing the faulty group within each module (Section~\ref{sec:fault_results}). Our analysis demonstrates that capacity and DVA-derived features separated the faulty cell groups within the modules with statistical significance, whereas internal resistance features did not for the modules in our dataset. However, the distributional overlap between faulty cell groups and non-faulty cell groups with respect to the health features highlights the challenges of utilizing these group-level electrochemical features for robust fault detection in the field. While we lack post-mortem physical analysis to confirm the specific root causes, the observed feature signatures allow us to hypothesize failure mechanisms consistent with the data. Our findings further the understanding of fault behavior in lithium-ion battery modules during field operation and, more importantly, demonstrate a framework for robustly monitoring the health of cells in lithium-ion battery modules under real-world conditions.

\section{Dataset}
\label{sec:dataset}

\subsection{Battery System}

% Writing Hardware configuration then faults
The dataset comprises operational data from 25 modules that were sourced from separate racks in one grid-connected BESS site. Each module consisted of 14 series-connected parallel groups, one of which was confirmed as faulty via post-mortem investigation, while the remaining 13 were assumed to be non-faulty. The fault type and the underlying cause were unknown. In total, the dataset included 25 faulty and 325 non-faulty cell groups. Of the 25 modules analyzed, 22 were configured as 14S3P and 3 were 14S2P.

% Writing Faults definition then Hardward configuration
% The dataset comprises operational data from 25 modules that were sourced from various, independent racks in one grid-connected BESS site. Of the 25 modules analyzed, 22 were configured as 14S3P and 3 were 14S2P. For each module, a single parallel-connected cell group was labeled as ``faulty" based on post-failure investigation, while the remaining 13 were assumed to be non-faulty. Fault type and the underlying cause of the fault were unknown. In total, the dataset included 25 faulty and 325 non-faulty cell groups.

Data were logged nominally at 1 Hz over a period of 1 to 6 months leading up to the day of the fault, generating over 183 million data rows (averaging 7.3 million data rows per module) and a total of 52,150 days of operational data across all cell groups. While the system targeted a 1 Hz sampling rate, the dataset contains gaps in recording, primarily during the post-discharge relaxation periods where the system likely entered a standby or data-saving mode. For each module, voltage was measured for every parallel-connected cell group with 1 mV precision. Current was measured at the module level. Since the 14 parallel-connected cell groups were arranged in series, this module-level current corresponds to the current flowing through each group in a module. Each module was equipped with only two temperature sensors. Consequently, there is no dedicated temperature measurement for each of the 14 series-connected cell groups; the sparse sensor coverage captures the overall thermal behavior of the module but limits the observability of the specific thermal states of individual cell groups.

The cells used were of an NMC/Graphite chemistry with a nominal capacity of 63 Ah and an operating voltage range of 3.0 to 4.2 V. 

Due to confidentiality agreements, the source of the BESS data cannot be disclosed.

\subsection{Field Operations}

\begin{figure}[h!]
    \centering
    \includegraphics[width=1\linewidth]{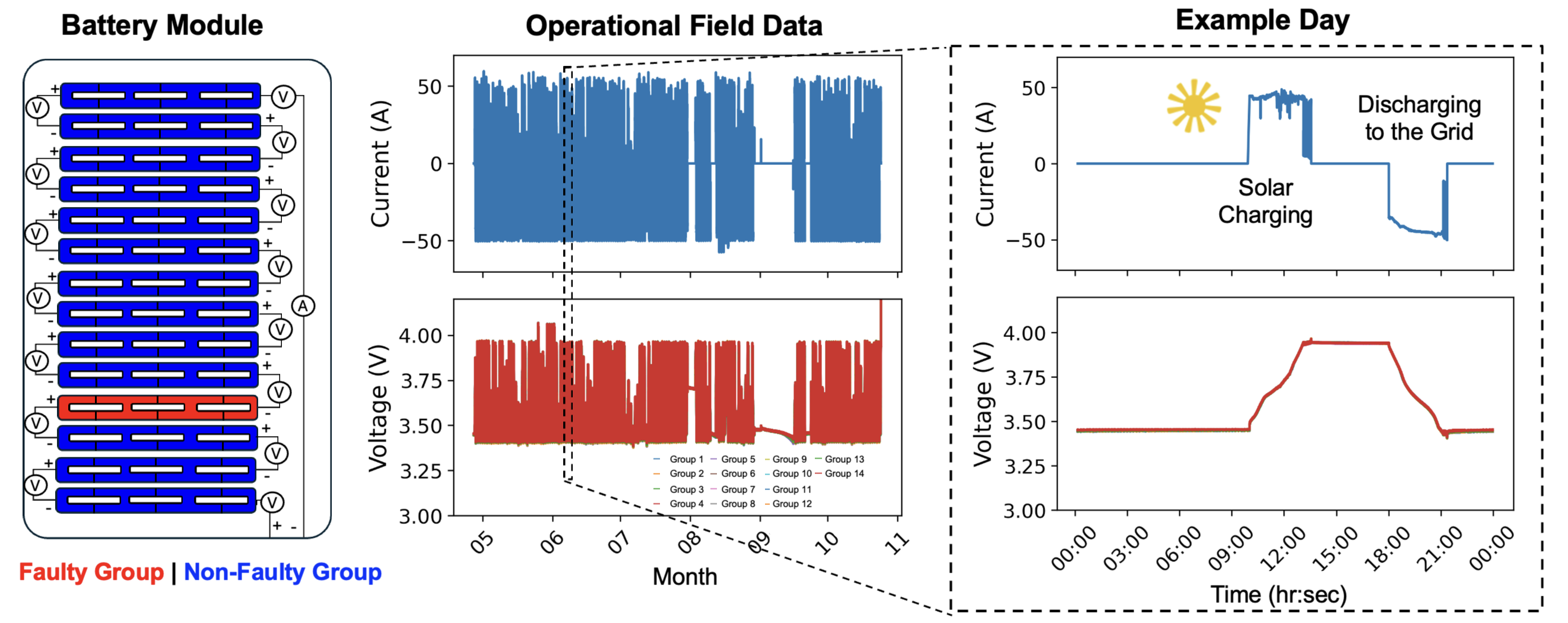}
    \caption[Example Data of a Grid-Connected BESS Module]{Example data of a single battery module. Typical daily usage patterns include 1) solar charging, 2) post-charge relaxation, 3) discharging to the grid, and 4) post-discharge relaxation.}
    \label{fig:figure_1_v2.png}
\end{figure}

\begin{figure}[h!]
    \centering
    \includegraphics[width=1\linewidth]{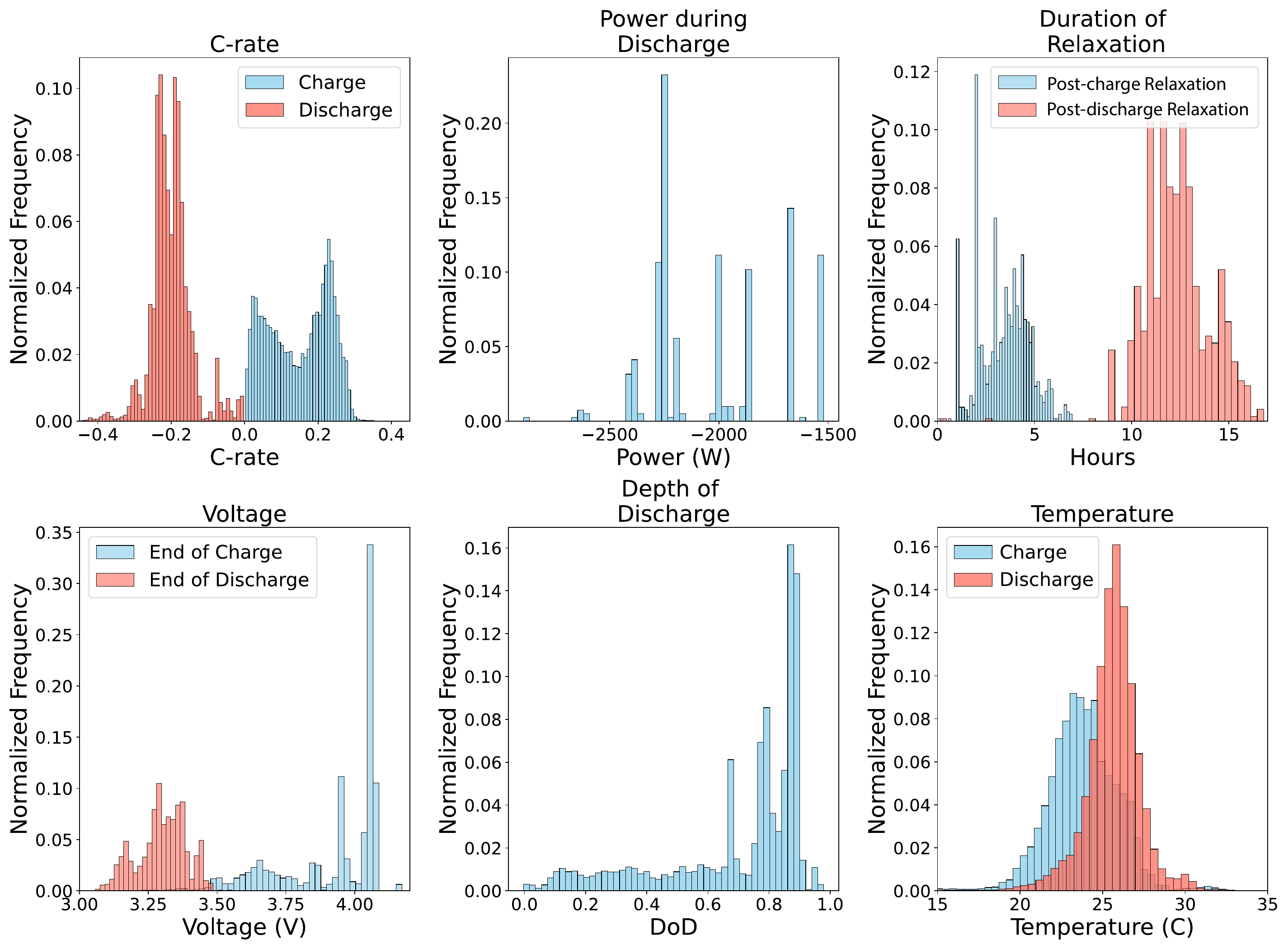}
    \caption[Distributions of Key Operational Parameters Across All Cell Groups and Cycles]{Distributions of key operational parameters across all cycles for the cell groups in the 25 modules. These histograms reveal the module-to-module and cycle-to-cycle variability in C-rate, discharge power, relaxation times, voltage limits, Depth of Discharge (DoD), and temperature.}
    \label{fig:all_operating_conditions_fault_modules_v3}
\end{figure}

Figure \ref{fig:figure_1_v2.png} displays a typical 24-hour operational cycle for each BESS module consisting of four distinct phases: solar charging, post-charge relaxation, discharging to the grid, and post-discharge relaxation. Figure \ref{fig:all_operating_conditions_fault_modules_v3} displays normalized histograms of key operational conditions experienced by the cell groups in the 25 modules across their days of cycling. During the day, modules underwent solar charging at a variable rate averaging 0.17 C until reaching a predefined rack-level State of Charge (SOC) limit, typically 80\% or 90\%. This was followed by a relaxation period averaging 3.45 hrs (std: 1.3 hrs). 
Subsequently, the modules discharged to the grid at a constant power until the rack-level SOC fell below 5\%. The discharge power level varied across modules, ranging from 1520 W to 2895 W. The discharge was followed by a final relaxation period lasting an average of 12.4 hrs (std: 1.6 hrs) until the next day's charging cycle. 

The operational data show that the cell groups within the modules were not under harsh operational stress. The distribution of the C-rate (defined as the current divided by the nominal Ampere-hour capacity of the battery) shows that the cell groups were not discharged or charged at a high current magnitude ($|I|< 0.45 C$). Furthermore, the end-of-charge and end-of-discharge voltage distributions show that cell groups did not overcharge or undercharge outside of the manufacturer-specified voltage limits (3.0 V to 4.2 V). Additionally, the temperature data indicates effective thermal management throughout the year, with the modules consistently maintained within a safe operating range. This suggests that the faulty cell groups within the modules did not arise from electrical or thermal abuse but rather likely from intrinsic degradation mechanisms.

However, the inherent variability in operating conditions, as shown in Figure~\ref{fig:all_operating_conditions_fault_modules_v3}, presents a significant challenge for extracting health features for cells. Fluctuations in conditions, such as C-rate and voltage windows, directly affect health features, confounding the true degradation trend with operational effects. To overcome this, it is essential to decouple the effects of operating conditions from the underlying signs of degradation. Accordingly, Section~\ref{sec:feature_sensitivity_to_operation_condition} introduces our GP regression framework, which calibrates the health features to isolate this time-dependent degradation trajectory.

\section{Data Processing and Feature Extraction}
\label{sec:feature_extraction}

Given that the fault type and the underlying cause were unknown, we developed algorithms to extract various health features from the cycling data to investigate which electrochemical properties might serve as effective indicators for detecting faulty cell groups. We focused on estimating capacity, internal resistances, and features derived from differential voltage analysis, as detailed in the following sections.

\subsection{Capacity}

\subsubsection{Method}
\label{subsec:cap_estimation_method}

\begin{figure}[h!]
    \centering
    \includegraphics[width=.9\linewidth]{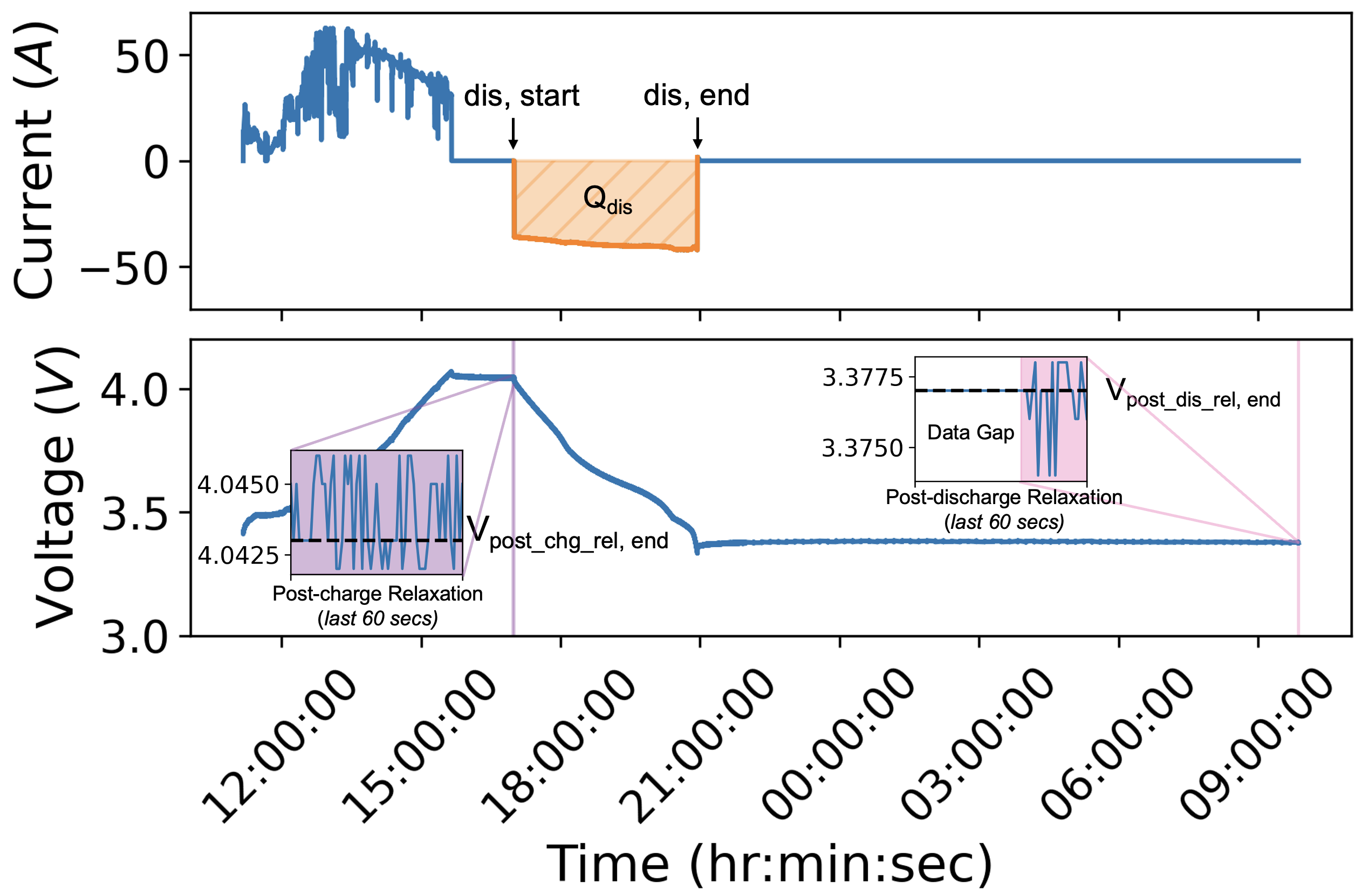}
    \caption[Methodology for Capacity Estimation from Operational Data]{Methodology for estimating the capacity of an exemplary cell group in a module using daily operational data.}
    \label{fig:capacity_estimation_v7}

\end{figure}

Capacity was estimated for each parallel-connected cell group during constant-power discharge events (Figure \ref{fig:capacity_estimation_v7}). We utilized constant-power discharge intervals for this estimation because their low-dynamic currents minimize numerical sampling and integration errors, enabling accurate Coulomb counting~\cite{en14144074}. Furthermore, the discharge events operated under consistent constant-power setpoints across cycles; this consistency minimizes the variability in capacity estimates induced by rate-dependence~\cite{Figgener2024}. The estimated cell-group capacity, $\widehat{C}$, was computed by dividing the Coulomb-counted discharge capacity, $Q_{\mathrm{dis}}$, by the change in state of charge (SOC) over the discharge interval,
$\widehat{z}_{\mathrm{dis},\,\mathrm{start}} - \widehat{z}_{\mathrm{dis},\,\mathrm{end}}$
(Equation~\ref{eq:capacity_estimate}). 

% In contrast, highly variable current during solar charging would amplify sampling and integration errors~\cite{en14144074}.

The SOC values at the start and end of discharge, $\widehat{z}_{\mathrm{dis},\,\mathrm{start}}$ and $\widehat{z}_{\mathrm{dis},\,\mathrm{end}}$, were determined by mapping the corresponding estimated open-circuit voltages to SOC using the OCV--SOC curve of a fresh cell (Equations~\ref{eq:z_start}--\ref{eq:z_end}; see ~\ref{subsec:fresh_OCV} for $\mathrm{OCV}_{\mathrm{fresh}}(z)$). These estimated OCVs were approximated using the terminal voltage measurements at the end of the post-charge and post-discharge relaxation periods, ${V}_{\mathrm{post\_chg\_rel},\,\mathrm{end}}$ and ${V}_{\mathrm{post\_dis\_rel},\,\mathrm{end}}$, respectively.
\begin{equation}
\widehat{C}
=
\frac{Q_{\mathrm{dis}}}
{\widehat{z}_{\mathrm{dis},\,\mathrm{start}}-\widehat{z}_{\mathrm{dis},\,\mathrm{end}}}
\label{eq:capacity_estimate}
\end{equation}

\begin{equation}
Q_{\mathrm{dis}}
=
-\frac{1}{3600}
\sum_{k = k_{\mathrm{dis},\,\mathrm{start}}}^{k_{\mathrm{dis},\,\mathrm{end}}-1}
\frac{I_k + I_{k+1}}{2}\,\bigl(t_{k+1}-t_k\bigr)
\label{eq:discharge_capacity}
\end{equation}

\begin{equation}
\widehat{z}_{\mathrm{dis},\,\mathrm{start}}
=
\mathrm{OCV}_{\mathrm{fresh}}^{-1}\!\left(
{V}_{\mathrm{post\_chg\_rel},\,\mathrm{end}}
\right)
\label{eq:z_start}
\end{equation}

\begin{equation}
\widehat{z}_{\mathrm{dis},\,\mathrm{end}}
=
\mathrm{OCV}_{\mathrm{fresh}}^{-1}\!\left(
{V}_{\mathrm{post\_dis\_rel},\,\mathrm{end}}
\right)
\label{eq:z_end}
\end{equation}

To reduce measurement noise and mitigate the influence of outliers, the voltage measurements were median-filtered over the final $60~\mathrm{s}$ of their respective relaxation periods, yielding robust OCV estimates over the quasi-static voltage windows (Equations~\ref{eq:ocv_dis_start}--\ref{eq:ocv_dis_end}).

\begin{equation}
{V}_{\mathrm{post\_chg\_rel},\,\mathrm{end}}
=
\operatorname{median}\!\left(
V(t)\;:\;
t \in [t_{\mathrm{\mathrm{post\_chg\_rel}},\,\mathrm{end}}-60,\; t_{\mathrm{\mathrm{post\_chg\_rel}},\,\mathrm{end}}]
\right)
\label{eq:ocv_dis_start}
\end{equation}
\begin{equation}
{V}_{\mathrm{post\_dis\_rel},\,\mathrm{end}}
=
\operatorname{median}\!\left(
V(t)\;:\;
t \in [t_{\mathrm{\mathrm{post\_dis\_rel}},\,\mathrm{end}}-60,\; t_{\mathrm{\mathrm{post\_dis\_rel}},\,\mathrm{end}}]
\right)
\label{eq:ocv_dis_end}
\end{equation}

A key assumption in this method is that the OCV-SOC relationship of the aged cell groups can be represented by that of the fresh cells. It is known that the OCV-SOC curve evolves with degradation due to aging mechanisms like loss of active material, and in turn can affect capacity estimation \cite{Zhou2025}. However, since this work prioritizes the reliable relative comparison of capacity among cell groups within a module over the absolute accuracy of any single capacity value, this assumption is justifiable. Applying the same OCV-SOC curve to all cell groups provides a consistent basis for this relative comparison.

\subsubsection{Implementation Details}

To ensure robust capacity estimation, we restricted our analysis to discharge cycles where ${V}_{\mathrm{post\_chg\_rel},\,\mathrm{end}} > 3.9~\mathrm{V}$ and ${V}_{\mathrm{post\_dis\_rel},\,\mathrm{end}}< 3.45~\mathrm{V}$.
This criterion was selected to minimize the propagation of voltage measurement noise into SOC estimation errors and, consequently, capacity estimation errors~\cite{Mohtat_2017}.
Voltages below $3.45~\mathrm{V}$ correspond to the steep region of the OCV curve, where the OCV gradient is high, thereby minimizing the sensitivity of $\widehat{z}_{\mathrm{dis},\,\mathrm{end}}$ to voltage measurement error.
Similarly, the region above $3.9~\mathrm{V}$ avoids the voltage plateau associated with the graphite Stage 2 phase transition ($\approx 3.7$--$3.9~\mathrm{V}$).
In this plateau region, the OCV gradient is shallow, causing small voltage errors to translate into significant variance in $\widehat{z}_{\mathrm{dis},\,\mathrm{start}}$ and, consequently, $\widehat{C}$.
By restricting ${V}_{\mathrm{post\_chg\_rel},\,\mathrm{end}}> 3.9~\mathrm{V}$, we ensure that the start of discharge is determined in a region with a sufficiently steep slope to yield a robust estimation of $\widehat{z}_{\mathrm{dis},\,\mathrm{start}}$.

In addition, to ensure that the voltages used for SOC estimation accurately represented open-circuit conditions, we only selected data where the post-charge and post-discharge relaxation periods used to estimate ${V}_{\mathrm{post\_chg\_rel},\,\mathrm{end}}$ and ${V}_{\mathrm{post\_dis\_rel},\,\mathrm{end}}$ lasted at least $1~\mathrm{h}$. This constraint increases confidence that the terminal voltage had sufficiently relaxed toward its open-circuit value prior to mapping to SOC using the fresh OCV--SOC relationship.

While these constraints ensured robust SOC estimations, thereby enabling accurate capacity estimations, they limited the quantity of analyzable discharge cycles. Although more complex online SOC estimation algorithms could be employed to relax these constraints and increase data utilization, such algorithms can struggle to account for rapid changes in capacity and resistance that can occur with faults~\cite{XU2025115524}. Therefore, the method employed here prioritizes robustness for fault detection over maximizing data quantity.

\subsection{Resistances}

\subsubsection{Method}

Internal resistances were estimated from the voltage relaxation behavior following solar charging events. The termination of each charging event is marked by an abrupt change in current, and the resulting voltage response to this current step during the subsequent relaxation period was used to estimate internal resistances of each cell group.

The estimated resistance at time $\tau$ into the post-charge relaxation period,
$\widehat{R}_{\mathrm{\mathrm{post\_chg\_rel}},\,\tau}$, was defined as the ratio of the change in terminal voltage ($\Delta V_\tau$) to the corresponding change in current between the final instant of the charging period and time $\tau$ ($\Delta I_\tau$) (Equation~\ref{eq:resistance_def}).

\begin{equation}
\widehat{R}_{\mathrm{\mathrm{post\_chg\_rel}},\,\tau}
= \frac{\Delta V_\tau}{\Delta I_\tau} = 
\frac{
V_{\mathrm{\mathrm{post\_chg\_rel}},\,\tau}
-
V_{\mathrm{chg},\,\mathrm{end}}
}{
I_{\mathrm{\mathrm{post\_chg\_rel}},\,\tau}
-
I_{\mathrm{chg},\,\mathrm{end}}
}
\label{eq:resistance_def}
\end{equation}

Here,
$V_{\mathrm{chg},\,\mathrm{end}}$ and
$I_{\mathrm{chg},\,\mathrm{end}}$
denote the terminal voltage and current at the final instant of the charging period, respectively, while
$V_{\mathrm{\mathrm{post\_chg\_rel}},\,\tau}$ and
$I_{\mathrm{\mathrm{post\_chg\_rel}},\,\tau}$
denote the corresponding quantities measured time $\tau$ into the post-charge relaxation period.

\begin{figure}[h!]
    \centering
    \includegraphics[width=1\linewidth]{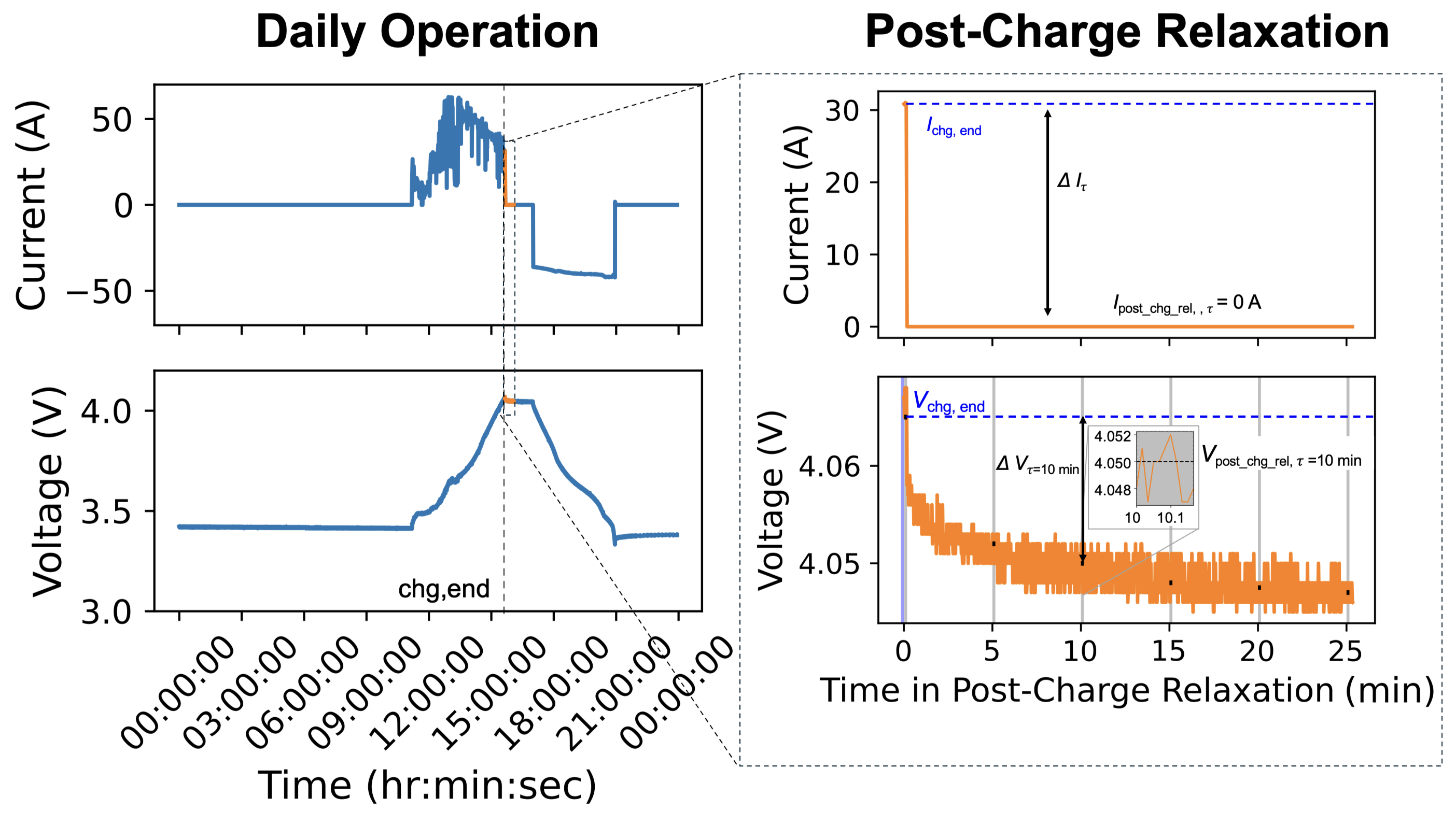}
    \caption[Methodology for Estimation of Resistances from Operational Data]{Methodology for estimating internal resistance of an exemplary cell group in a module using daily operational data.}
    \label{fig:resistance_estimations_v6}
\end{figure}

To reduce measurement noise and mitigate the influence of outliers, voltage measurements used in the resistance calculation were median-filtered over short time windows. Specifically, we defined
\begin{align}
V_{\mathrm{chg},\,\mathrm{end}}
&=
\operatorname{median}\!\left(
V(t)\;:\;
t \in [t_{\mathrm{chg},\,\mathrm{end}}-\Delta t_{\mathrm{chg}},\; t_{\mathrm{chg},\,\mathrm{end}}]
\right),
\label{eq:V_chg_end_med}
\\
V_{\mathrm{\mathrm{post\_chg\_rel}},\,\tau}
&=
\operatorname{median}\!\left(
V(t)\;:\;
t \in [t_{\mathrm{\mathrm{post\_chg\_rel}},\,\mathrm{start}}+\tau,\; t_{\mathrm{\mathrm{post\_chg\_rel}},\,\mathrm{start}}+\tau+\Delta t_{\mathrm{rel}}]
\right),
\label{eq:V_rel_chg_x_med}
\end{align}
where $\Delta t_{\mathrm{chg}}$ and $\Delta t_{\mathrm{rel}}$ denote the median-filter window lengths (in seconds). We set $\Delta t_{\mathrm{chg}}=\Delta t_{\mathrm{rel}}=10~\mathrm{s}$, which at 1~Hz provides a robust median over $\sim$10 samples while remaining short relative to the relaxation timescale. Current measurements were used without filtering.

We extracted resistance estimates with $\tau \in \{1\,\text{s}, 5\,\text{min}, 10\,\text{min}, 15\,\text{min}, 20\,\text{min}, 25\,\text{min}\}$. Resistance estimates obtained at different values of $\tau$ capture distinct physical contributions. The resistance evaluated shortly after the current interruption,
$\widehat{R}_{\mathrm{\mathrm{post\_chg\_rel}},\,1~\mathrm{s}}$,
primarily reflects the ohmic resistance of the cell group, while resistance estimates obtained further into the relaxation period ($\tau > 1~\mathrm{s}$) increasingly incorporate charge-transfer and diffusion-related resistances.

\subsubsection{Implementation Details}
\label{subsec:implementation_details_resistance}

To ensure reliable resistance estimates, the absolute change in current
($|\Delta I_\tau|$), was required to exceed a minimum amplitude of $20~\mathrm{A}$. This threshold ensures that the corresponding absolute voltage change ($|\Delta V_\tau|$) is significantly larger than the voltage sensor resolution ($1~\mathrm{mV}$), thereby providing a sufficiently high signal-to-noise ratio for reliable resistance estimation.

\subsection{Differential Voltage Analysis (DVA) Features}

\subsubsection{Method}
\begin{figure}[h!]
    \centering
    \includegraphics[width=1\linewidth]{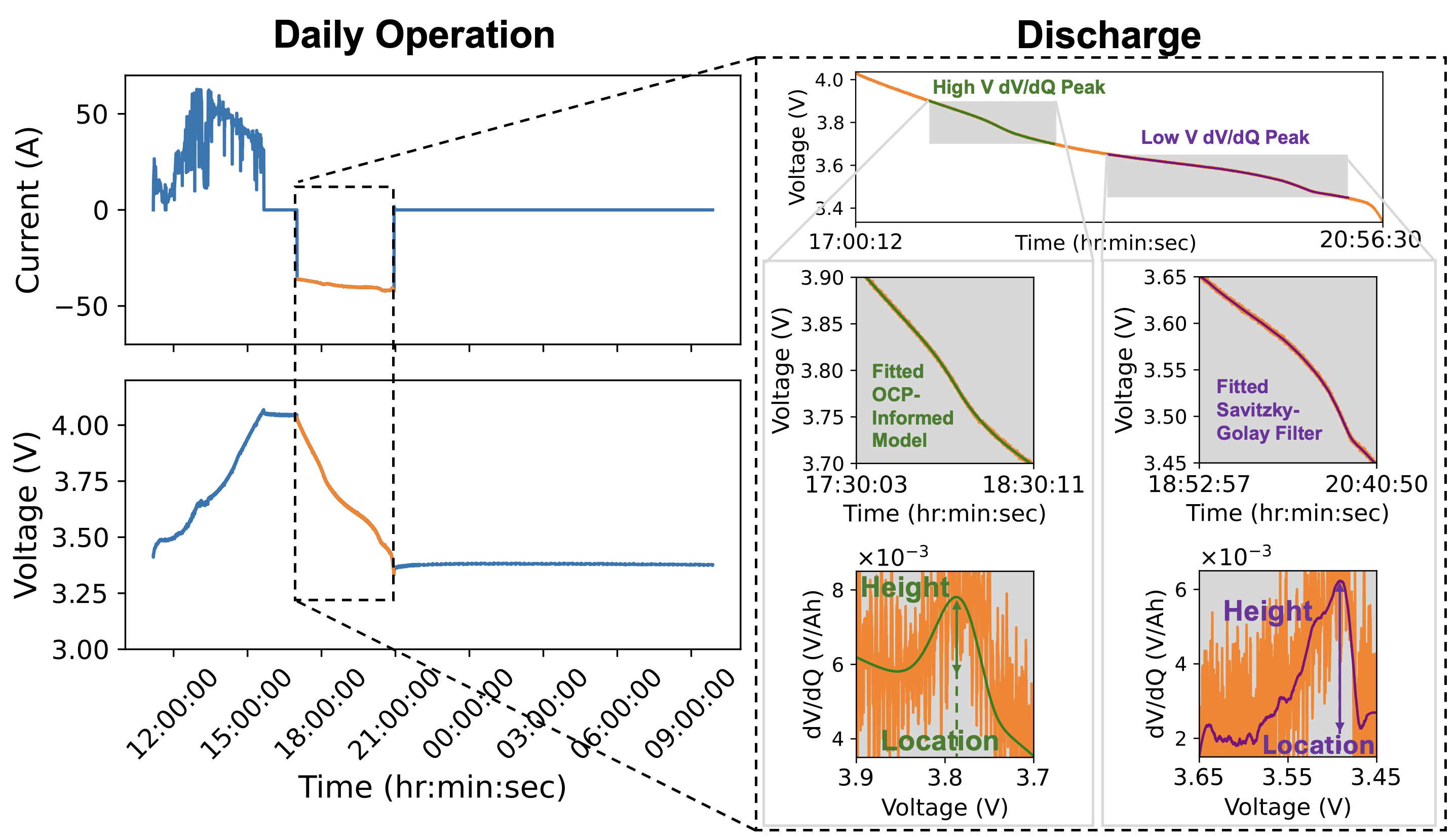}
    \caption[Methodology for $dV/dQ$ Feature Extraction from Operational Data]{Methodology for $dV/dQ$ feature extraction  of an exemplary cell group in a module using daily operational data.
    
    }
    \label{fig:dVdz_estimations_v4}
\end{figure}

Differential voltage analysis features were estimated for each cell group using the constant power discharge portions of the field data. These features correspond to the peaks that arise in the dV/dQ curve of the cell groups under low-dynamic discharge. For our NMC/graphite cell groups, two peaks in the dV/dQ curve can appear: High V dV/dQ peak, which occurs between 3.7 V and 3.9 V, and Low V dV/dQ peak, which occurs between 3.45 V to 3.65 V. Four features are extracted from these curves: the height and voltage location of each of these two peaks (Figure \ref{fig:dVdz_estimations_v4}).

However, accurately identifying these features is challenging due to measurement noise from the inexpensive sensors used in the field~\cite{SULZER20211934, Antti_Field}. To address this, we implemented specific strategies for each peak. 

For High V dV/dQ peak, we applied the \textit{OCP-informed Feature Identification} algorithm presented in \cite{Wong_2024}. This algorithm fits the voltage and charge throughput data corresponding to High V dV/dQ peak to a physics-based model to determine the peak’s height and location. A key advantage of this approach is that it avoids the need for user-defined tuning parameters required by traditional filtering techniques, thereby enabling accurate and reproducible feature estimation.

The \textit{OCP-informed Feature Identification} algorithm, however, is only applicable to well-defined, isolated peaks and thus cannot be applied to Low V dV/dQ peak, as multiple phase transitions occur within this voltage window, causing a broad and often distorted peak. For Low V dV/dQ peak, we first applied a Savitzky-Golay (SG) filter to the voltage data. For the SG filter, we selected a third-order polynomial, which effectively models the graphite phase transition without overfitting to noise. We set the window size to span a number of data points equivalent to a 7\% SOC window, which makes the peak clearly identifiable. Having filtering parameters set to a fixed polynomial order and fixed SOC window led to consistent filtering \cite{Mohtat2022}. After smoothing, we applied a peak-finding algorithm to identify Low V dV/dQ peak and valley. The peak height was then computed as the difference between the peak and valley, and its location was recorded as the voltage of the peak.

\subsubsection{Implementation Details}
To ensure the integrity of the dV/dQ features, we excluded discharge segments where the power level changed within the voltage window of the specific peak being analyzed.

\section{Gaussian Process (GP) Regression for Feature Calibration and Uncertainty Quantification}
\label{sec:feature_sensitivity_to_operation_condition}

The health features extracted from operational cycling data in Section \ref{sec:feature_extraction} exhibited high sensitivity to cycle-to-cycle changes in operating conditions, which complicated isolating the underlying degradation trend over time. Table \ref{health_features_table} summarizes the health features for each cell group and the corresponding operating conditions to which each feature is highly sensitive.

\begin{table}[h!]
    \centering
    \caption[Summary of Health Features and the Operating Conditions They Are Sensitive to]{Summary of health features ($y$) and the operating conditions ($\mathbf{x}$) they are sensitive to}
    \label{health_features_table}
    \begin{tabular}{
        >{\centering\arraybackslash}m{0.4\linewidth}
        >{\centering\arraybackslash}m{0.5\linewidth}
        }
        \toprule
        \textbf{Health Feature ($y$)} & \textbf{Operational Inputs ($\mathbf{x}$)} \\
        \midrule
        DVA Peak Heights \& Locations & Voltage at start of discharge ($V_{\mathrm{dis},\,\mathrm{start}}$),  C-rate \\
        \midrule
        Resistance ($\widehat{R}_{\mathrm{\mathrm{post\_chg\_rel}},\,\tau}$) & Current ($|\Delta I_\tau|$), SOC \\
        \midrule
        Capacity ($\widehat{C}$) & C-rate, Voltage at end of post-charge relaxation (${V}_{\mathrm{post\_chg\_rel},\,\mathrm{end}}$)  \\
        \bottomrule
    \end{tabular}
\end{table}

To address these sensitivities to operating conditions, we employed a GP regression framework to calibrate each health feature for each cell group. We selected a GP framework for this task due to several key advantages. First, GP models are non-parametric models that can capture complex, nonlinear relationships without a fixed functional form and learn dependencies directly from the data \cite{GP_textbook}. Second, GP models provide a principled measure of uncertainty for each calibrated feature \cite{GP_textbook}, which is critical for distinguishing whether a feature's value reflects abnormal behavior or is merely a byproduct of operating conditions and sensor noise.

We modeled each cell group's health feature ($y$) as a GP model ($f$) composed as the sum of two independent GPs: $f_{op}$, which captures the feature's sensitivity to operating conditions ($x$), and $f_{deg}$, which captures how the feature changes with time ($t$) which is indicative of degradation. This additive structure allows the model to decouple the operational dynamics while simultaneously learning the feature's sensitivity to operating conditions. However, this structure assumes that the relationship between the features and time is independent of the features' sensitivity to operating conditions. Equations \ref{eq: GP_overview_1} - \ref{eq: GP_overview_4} overview the modeling equations for a given cell group, with independent Gaussian noise \(\epsilon\):
\begin{equation}
\label{eq: GP_overview_1}
y(\mathbf{x}, t) = f(\mathbf{x}, t) + \epsilon, \quad \text{where} \quad \epsilon \sim \mathcal{N}(0, \sigma_n^2)
\end{equation}
\begin{equation}
\label{eq:GP_model}
f(\mathbf{x}, t) = f_{op}(\mathbf{x}) + f_{deg}(t)
\end{equation}
\begin{equation}
f_{op}(\mathbf{x}) \sim \mathcal{GP}(\mu_{op}(x), k_{op}(\mathbf{x}, \mathbf{x}'))
\end{equation}
\begin{equation}
\label{eq: GP_overview_4}
f_{deg}(t) \sim \mathcal{GP}(\mu_{deg}(t), k_{deg}(t, t'))
\end{equation}

We select the kernels for $f_{op}$ and $f_{deg}$ based on physical assumptions about the batteries' behavior. For modeling the dependency on operating conditions, we used a radial basis function (RBF) kernel, assuming each feature's response to changing operating conditions is relatively smooth. For modeling the dependency on time, we used a non-stationary Brownian motion kernel, since its cumulative properties provide a direct mathematical analogue for the irreversible, accumulating process of battery degradation. The full details about kernels, implementation, and hyperparameter optimization are provided in ~\ref{sec:GP_modeling_framwork}.

With our GP method decoupling sensitivities to operating conditions, our method enables continuous health monitoring whenever features are extractable, regardless of the specific operating conditions. For instance, across the 52,150 days of operational data in the dataset, the High V dV/dQ peak height is extractable on 30,492 days. Without our method, ensuring that changes in the High V dV/dQ peak height reflect degradation rather than operational variability would require downselecting data for each cell group to the most common C-rate and voltage at the start of discharge, the operating conditions to which the High V dV/dQ peak height feature is sensitive. This downselection results in only 17,771 usable High V dV/dQ peak height values across the dataset. Our GP method enables the utilization of all 30,492 extractable values.

Sections \ref{subsec:DVA_operating_conditions} - \ref{subsec:cap_operating_conditions} present the sensitivities of the health features to operating conditions learned from the GP models, which are consistent with battery physics and validate the GP models' physical plausibility.

\subsection{Sensitivity of DVA features to operating conditions}
\label{subsec:DVA_operating_conditions}
\begin{figure} [h!]
    \centering
    \includegraphics[width=.85
    \linewidth]{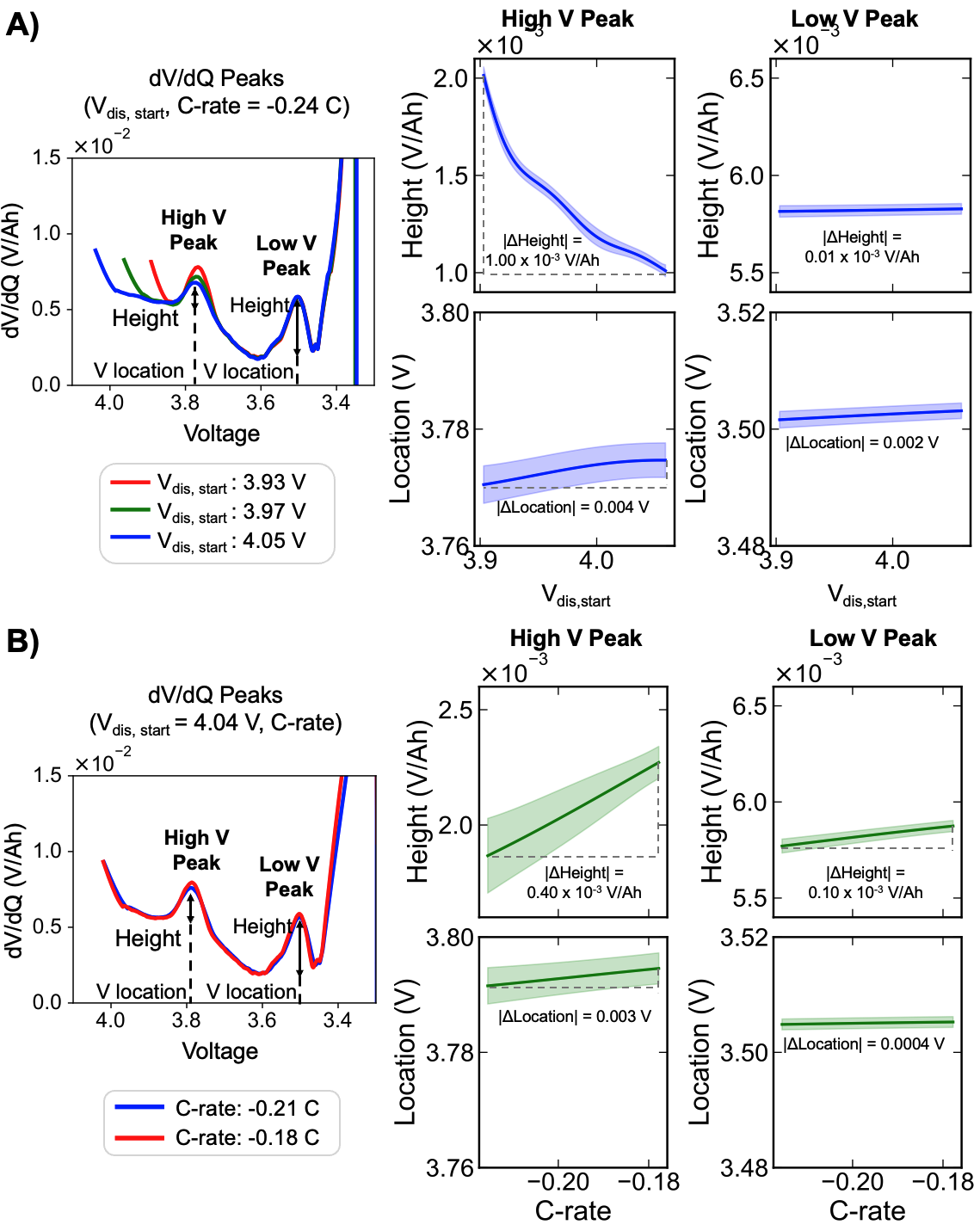}
    \caption[GP-Learned Sensitivities of DVA Features to Voltage at Start of Discharge and C-Rate]{GP-learned sensitivities of DVA features to $V_{\mathrm{dis},\,\mathrm{start}}$ and C-rate. Plots illustrate the mean (solid line) and 95\% confidence interval (shaded area) of the learned relationships. 
    (A) Sensitivity to $V_{\mathrm{dis},\,\mathrm{start}}$, learned from Cell 1 in Module 23, with C-rate fixed at its median. (B) Sensitivity to C-rate, learned from Cell 1 in Module 6, with $V_{\mathrm{dis},\,\mathrm{start}}$ fixed at its median. Cell 1 in Module 23 and 6 were selected for exhibiting the widest variance in $V_{\mathrm{dis},\,\mathrm{start}}$ and C-rate, respectively}
    \label{fig:DVA_features_to_operating_conditions}
\end{figure}

The differential voltage analysis features depend on the voltage at the start of discharge ($V_{\mathrm{dis},\,\mathrm{start}}$) and the C-rate of the power discharge~\cite{Mohtat2022, Weng2023-lj}.
Figure \ref{fig:DVA_features_to_operating_conditions} illustrates these dependencies, as learned by the GP models. The plots are generated from models trained on cell groups selected for their wide variation in the respective operating conditions, as these cases best highlight the learned relationships.

$V_{\mathrm{dis},\,\mathrm{start}}$ primarily affects the High V peak features, particularly its height, while the Low V peak features show negligible sensitivity. With increasing $V_{\mathrm{dis},\,\mathrm{start}}$, High V peak height significantly decreases. The learned relationship directly aligns with lithium-ion diffusion dynamics. The High V peak appears early in the discharge before diffusion processes have fully stabilized. $V_{\mathrm{dis},\,\mathrm{start}}$ dictates the time required to reach this peak's associated phase transition, which in turn affects the cell group's diffusion gradients at the time of the peak and consequently the peak's features. Conversely, the Low V peak features are insensitive to $V_{\mathrm{dis},\,\mathrm{start}}$ because this peak occurs later in the discharge, by which point the initial diffusion dynamics have stabilized \cite{Mohtat2022}.

In contrast to $V_{\mathrm{dis},\,\mathrm{start}}$, C-rate affects the features of both the High V and Low V peaks. The peak heights are particularly sensitive, decreasing significantly with higher C-rates, while the peak locations minimally change within the range of C-rate. This trend is consistent with battery physics: higher C-rates increase cell polarization and kinetic limitations, which broaden the dV/dQ peaks, thus reducing their height and shifting their location \cite{Weng2023-lj, Mohtat_2020}.

\subsection{Sensitivity of Resistances to Operating Conditions}
\label{subsec:R_operating_conditions}

Internal resistances depend on a battery's age, current, temperature, and SOC \cite{Antti_Field, Schaeffer2024}. We excluded the temperature dependence from our  GP modeling because the observed thermal variation was minimal (std:1.3 °C) and the available sensors do not measure the local temperature of each individual cell group.

\begin{figure} [h!]
    \centering
    \includegraphics[width=1
    \linewidth]{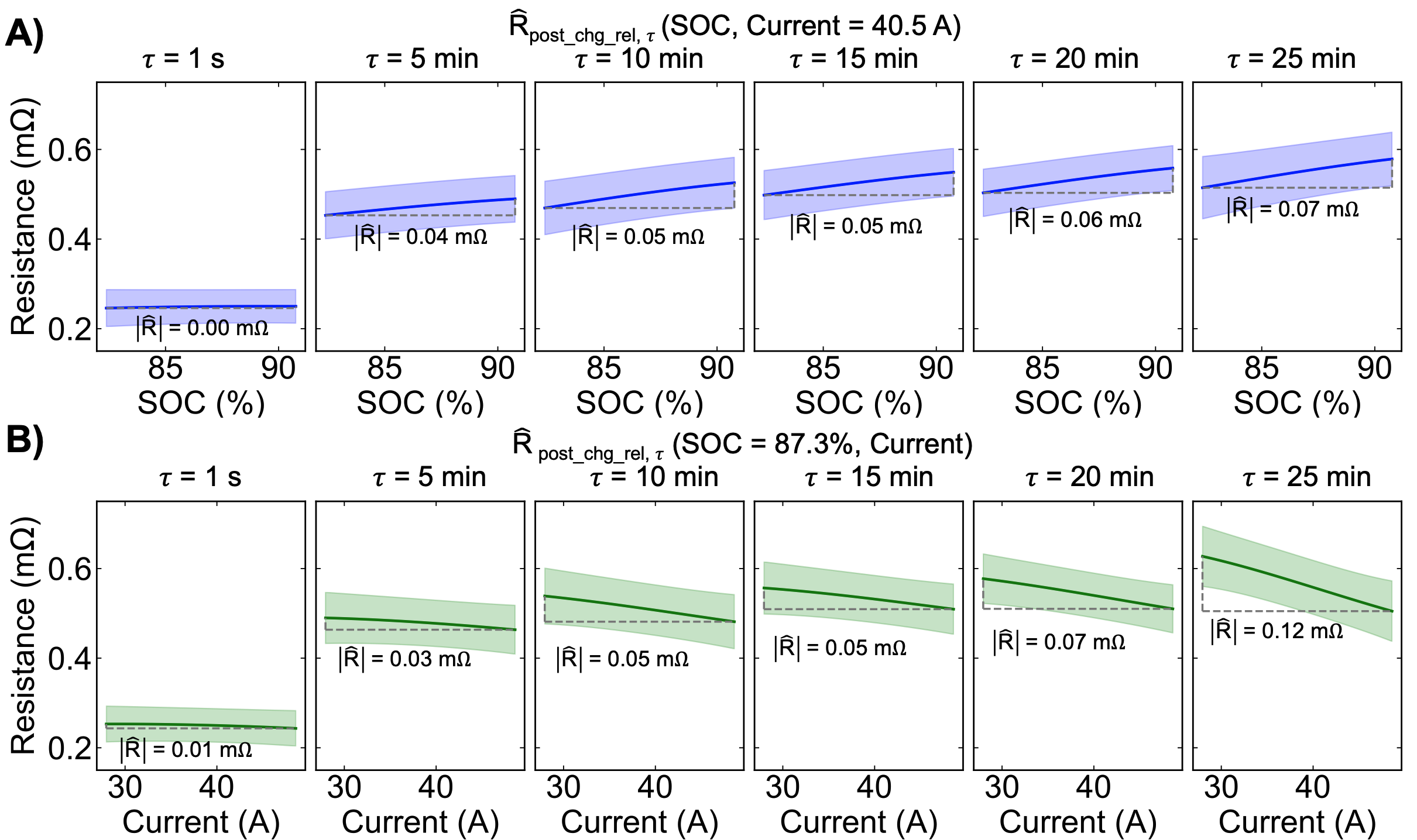}
    \caption[GP-Learned Sensitivities of Internal Resistances to SOC and Current]{GP-learned sensitivities of internal resistances to SOC and current. Plots illustrate the mean (solid line) and 95\% confidence interval (shaded area) of the learned relationships. (A) Sensitivity to SOC, learned from Cell 1 in Module 14, with current magnitude fixed at its median. (B) Sensitivity to current, learned from Cell 1 in Module 14, with SOC fixed at its median. Cell 1 in Module 14 was selected for exhibiting the widest variance in both SOC and current.}
    \label{fig:R_vs_operating_conditions}
\end{figure}

Figure~\ref{fig:R_vs_operating_conditions} shows the variability of the estimated internal resistances ($\widehat{R}_{\mathrm{\mathrm{post\_chg\_rel}},\,\tau}$) as a function of SOC ($\widehat{z}_{\mathrm{\mathrm{post\_chg\_rel}}}$) and current ($|\Delta I_\tau|$) learned from our GP modeling. As shown, the resistance values change significantly with operating conditions, and this sensitivity becomes more pronounced at longer relaxation times.

Resistances were observed to monotonically decrease as the current increases, which is consistent with findings in previous studies~\cite{Ansean2013, Antti_Field, Schaeffer2024}. This trend can be explained by electrochemical processes governed by Butler-Volmer kinetics, where the activation overpotential required for charge transfer increases more slowly at higher current densities (see \ref{appendix:resistance_derivation} for more details)~\cite{Bai2014}.

Resistances were observed to change with SOC in the high-SOC region where the cells typically operated, increasing slightly from 80\% to 90\% SOC. This is consistent with previous studies~\cite{Schaeffer2024, WENG_formation_prediction} and can be attributed to thermodynamic and mass-transport limitations at the graphite anode. As the graphite anode becomes more saturated at higher SOC, a greater overpotential is required to insert ions into the packed structure, which increases the charge-transfer resistance, while slow internal diffusion increases the mass-transport resistance \cite{Lin_resistance_operating_conditions}.

The increasing sensitivity of resistances to both SOC and current as the relaxation time ($\tau$) increases is consistent with battery physics, as these resistances corresponding to instances longer into relaxation ($\tau>1$ s) increasingly incorporate charge-transfer resistance and diffusion resistance. The estimated resistance at time $\tau=1$ s into the post-charge relaxation period  ($\widehat{R}_{\mathrm{\mathrm{post\_chg\_rel}},\,1\,\mathrm{s}}$) shows minimal sensitivity as it is dominated by ohmic resistance which is largely independent of SOC and current. By $\tau=5$ minutes, $\widehat{R}_{\mathrm{\mathrm{post\_chg\_rel}},\,5\,\mathrm{min}}$ becomes significantly more sensitive due to the inclusion of charge-transfer resistance, which is governed by nonlinear electrochemical kinetics and is highly dependent on both SOC and current. At $\tau\geq10$ minutes, these sensitivities are amplified further by the inclusion of diffusion resistance, which is also highly sensitive to SOC and current. Ultimately, this progression explains why $\widehat{R}_{\mathrm{\mathrm{post\_chg\_rel}},\,25\,\mathrm{min}}$ is the most sensitive to operating conditions \cite{Plett_textbook}.

\subsection{Sensitivity of Capacity to Operating Conditions}
\label{subsec:cap_operating_conditions}
\begin{figure} [h!]
    \centering
    \includegraphics[width=1
    \linewidth]{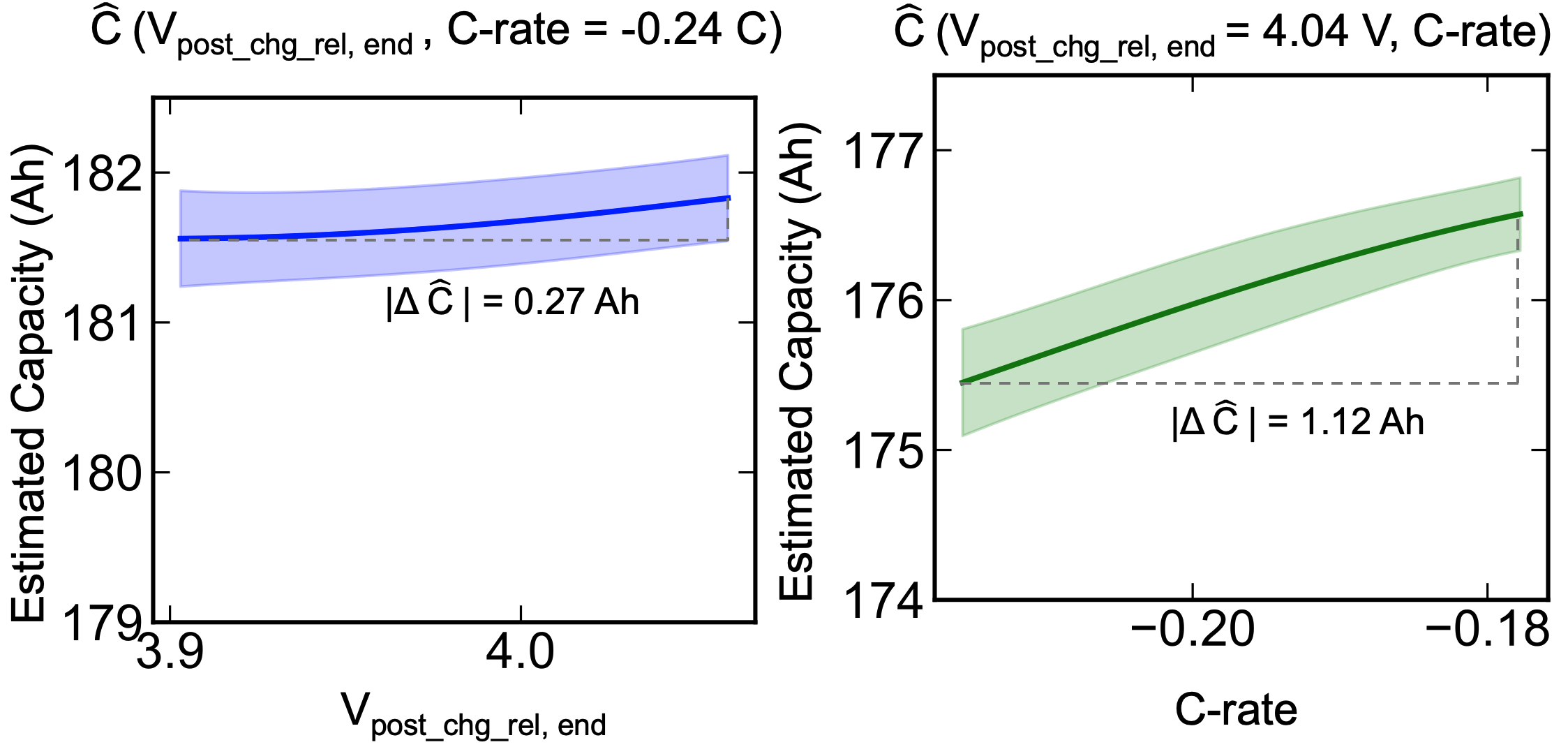}
    
    \caption[GP-Learned Sensitivities of Capacity Estimates to Voltage at the End of Post-charge Relaxation and C-Rate]
    {GP-learned sensitivities of capacity estimates to $V_{post\_chg\_rel,\,end}$ and C-rate. Plots illustrate the mean (solid line) and 95\% confidence interval (shaded area) of the learned relationships.(A) Sensitivity to $V_{post\_chg\_rel,\,end}$, learned from Cell 1 in Module 23, with C-rate fixed at its median. (B) Sensitivity to C-rate, learned from Cell 1 in Module 6, with $V_{post\_chg\_rel,\,end}$ fixed at its median. Cell 1 in Module 23 and 6 were selected for exhibiting the widest variance in $V_{post\_chg\_rel,\,end}$ and C-rate, respectively.}
    \label{fig:cap_vs_operating_conditions}
\end{figure}

Capacity estimations ($\widehat{C}$) depend on the operating conditions of the discharge segment, particularly the applied C-rate, the estimated voltage at the end of post-charge relaxation (${V}_{\mathrm{post\_chg\_rel},\,\mathrm{end}}$), and the estimated voltage at the end of post-discharge relaxation (${V}_{\mathrm{post\_dis\_rel},\,\mathrm{end}}$) \cite{HE2021102867, Figgener2024}. In our dataset, there was minimal variation in the estimated end-of-discharge SOC ($\widehat{z}_{\mathrm{dis},\,\mathrm{end}}$) for each cell in a module. Accordingly, we designed the operational component of our Gaussian process model for each cell group's capacity estimations to use only C-rate and ${V}_{\mathrm{post\_chg\_rel},\,\mathrm{end}}$ as inputs.

Figure \ref{fig:cap_vs_operating_conditions} shows these learned dependencies, which are physically consistent with established electrochemical principles. First, the model captures that higher magnitude C-rates lead to lower estimated capacity, which is attributed to increased voltage drops across the battery's internal resistance at higher currents causing the terminal voltage to reach its cutoff limit prematurely~\cite{Figgener2024}. Furthermore, the model captures the sensitivity of the capacity estimation to ${V}_{\mathrm{post\_chg\_rel},\,\mathrm{end}}$, which is a direct consequence of the battery's non-linear OCV curve. Due to the variable slope of this curve, constant voltage measurement errors translate into variable SOC uncertainties. A value of ${V}_{\mathrm{post\_chg\_rel},\,\mathrm{end}}$ located in a steeper OCV region reduces SOC error, effectively minimizing the overestimation of $\Delta \text{SOC}$ and yielding a higher capacity estimate (See \ref{subsec:supporting_work_for_cap_vs_vstart} for more details)~\cite{QI2024605}. In Figure \ref{fig:cap_vs_operating_conditions}, the capacity estimate at $4.05\,\mathrm{V}$ is higher than at $3.90\,\mathrm{V}$ due to the steeper OCV gradient at $4.05\,\mathrm{V}$ (Figure \ref{fig:dOCV}).

\subsection{Calibrated Health Features for Cell Groups}
\label{subsec:intra-module-feature-calibration}
\begin{figure}[h!]
    \centering
    \includegraphics[width = .8\linewidth]{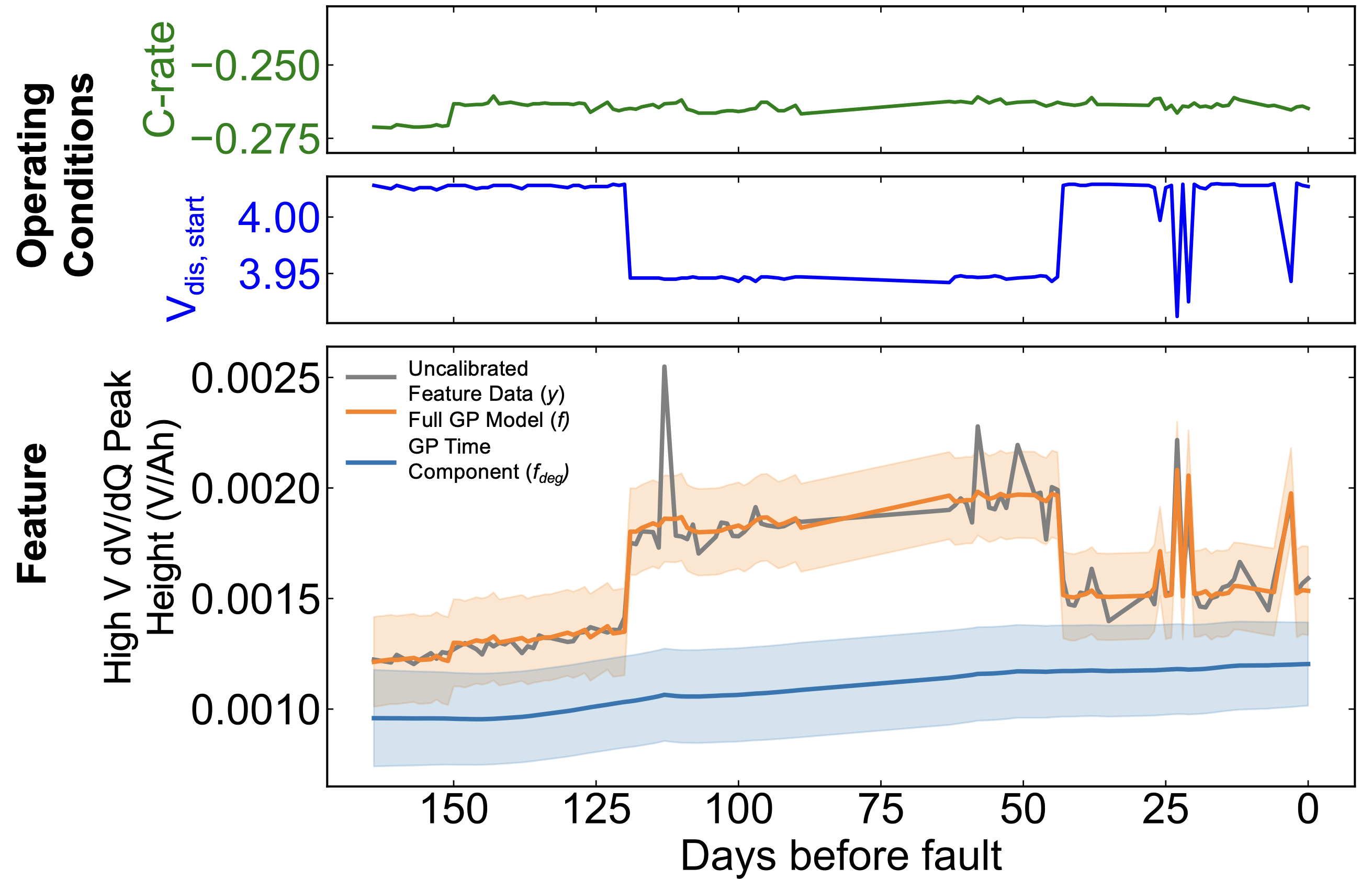}
    \caption[Visual Demonstration of Gaussian Process Calibration on a Health Feature]{Visual demonstration of Gaussian process calibration on a health feature (e.g. High V Peak Height). The uncalibrated feature data ($y$) is fitted to a full GP model (f) that accounts for fluctuations caused by operating conditions. After model fitting, the GP time component ($f_{deg}$) reveals the health feature's underlying aging trend. Shaded areas denote the 95\% confidence intervals. }
    \label{fig:GP learned data}
\end{figure}
Our GP modeling isolates each health feature's underlying aging trend ($f_{deg}$) and decouples the effects of operating conditions. Figure \ref{fig:GP learned data} provides a visual demonstration of this calibration process through GP modeling for an example health feature (e.g. High V dV/dQ Height) of a single cell group. The uncalibrated health feature data ($y$) shows the feature's variance due to changes in the operating conditions. The full GP model ($f$) is fitted to $y$. Once fitted, we compute the posterior mean of the time-dependent component  ($f_{deg}$) as the calibrated feature, effectively decoupling the degradation trend from the feature's sensitivity to operating conditions ($f_{op}$).

\begin{figure}[h!]
    \centering
    \includegraphics[width=\linewidth]{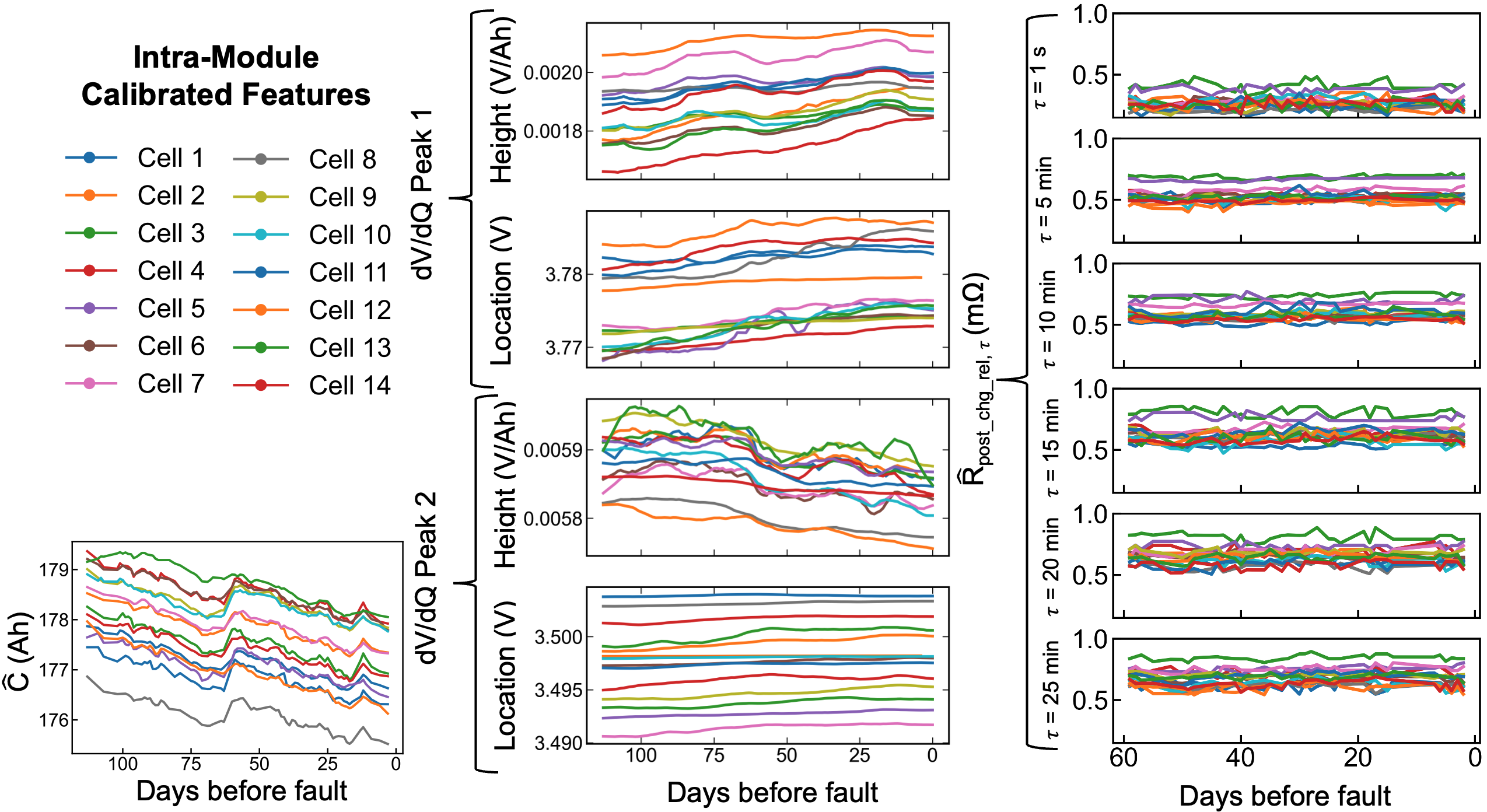}
    \caption[Illustration of Calibrated Health Features of Cell Groups in a Module]{Calibrated capacity, DVA, and resistances of cell groups in a module (all 14 cell groups in module 8). The calibration process removes effects from operational variability, revealing clearer and more monotonic aging trends.}
    \label{fig:Calibrating_DVA}
\end{figure}

Figure \ref{fig:Calibrating_DVA} shows the resulting calibrated trajectories for all health features and all cell groups within an example module. From these isolated trajectories, we can extract the feature's estimated value and its rate of change (derivative) at any point in time. Both features are potentially useful indicators for identifying faults in modules.

\section{Evaluating Fault Diagnostic Potential of Individual Features}
\label{sec:fault_results}

From the GP-calibrated time series of capacity, internal resistance, and DVA features (Section~\ref{subsec:intra-module-feature-calibration}), we extracted median values from varying pre-fault time windows (0–15, 15–30, and 30–45 days) and linear slopes calculated across all available cycles. The median values capture the cell group's absolute health features within a specific time window, while the slope captures the cell group's rate of degradation.

Before evaluating the fault diagnostic potential of these features, we normalized them relative to the cell groups within each module to ensure comparability across modules with different operating conditions (Section~\ref{subsec: normalization}). Using these normalized features, we then evaluated their fault diagnostic potential. First, we assessed each feature's ability to separate the faulty and non-faulty cell groups with statistical significance using Mann–Whitney U tests (Section \ref{subsec:p-value_results}). Second, we quantified the discriminative utility of each feature for fault detection across all possible decision thresholds (Section \ref{subsec:auc_results}).

\subsection{Normalizing Features for Cross-Module Feature Comparability}

\label{subsec: normalization}

The GP model (Equation \ref{eq:GP_model}; $f(\mathbf{x}, t) = f_{op}(\mathbf{x}) + f_{deg}(t)$) for each cell group's health feature is trained on that specific group's data. While the model effectively decouples the feature's sensitivity to cycle-to-cycle operational condition changes ($f_{op}(\mathbf{x})$), the values of the calibrated feature ($f_{deg}(t)$) are learned relative to the cell group's specific operating conditions ($x$).

Within a module, the operating conditions of cell groups are relatively similar. All cell groups within the same module share the same module-level current and are kept well-balanced in voltage by the Battery Management System (BMS) (\ref{subsec:Intra-Module Uniformity}). Because the cell groups within a module share these common operating conditions, differences in their calibrated health features reflect intrinsic differences in health rather than differences in operating conditions.

However, operating conditions differ for cell groups in different modules. Because a cell group's calibrated health features are learned relative to each cell group's specific operating conditions, their values are not directly comparable across modules. 

\begin{figure}[h!]
    \centering
    \includegraphics[width=1\linewidth]{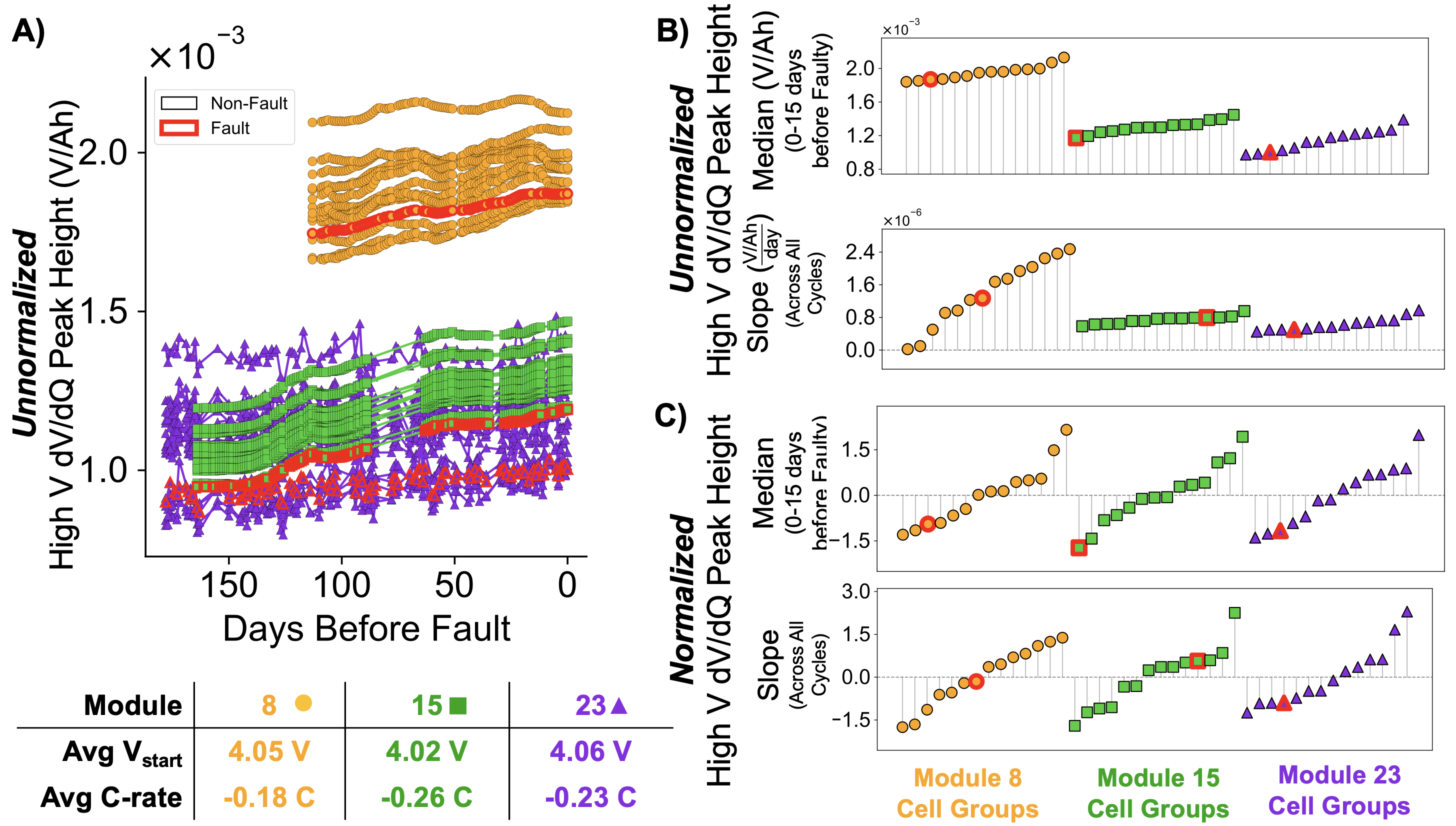}
    \caption[Feature Normalization Process for Enabling Cross-Module Comparability]{Feature normalization to enable cross-module comparability, using High V Peak Height as an example feature. (A) Unnormalized time-series data for three modules (8, 15, 23), showing that differences in operating conditions cause offsets in feature values across modules. (B) Unnormalized features (median (0--15 days before fault) and slope (across all cycles)) extracted from the time-series data. Because these values are affected by their module's specific operating conditions, they are unreliable for cross-module fault detection. (C) The same features after normalization.}
    \label{fig:normalization_example}
\end{figure}

Figure~\ref{fig:normalization_example}(A) illustrates how differences in operating conditions across modules cause offsets in the values of the High V dV/dQ Peak Height, used here as a representative example. These offsets propagate to derived features, such as the median value of the feature at a pre-fault time window (0--15 days) and the slope of the feature value across cycling, as shown in Figure~\ref{fig:normalization_example}(B)

To evaluate the effectiveness of a health feature to separate faulty cell groups, the feature values must be made comparable across modules. We achieve this by normalizing each cell group's feature value relative to the feature values of other cell groups within the same module, as defined in Equation~\ref{eq:feature_normalization}:

\begin{equation}
\label{eq:feature_normalization}
\begin{aligned}
\phi_i^{\mathrm{norm}} &= \frac{\phi_i - \mu_{\mathrm{module}}}{\sigma_{\mathrm{module}}}, \\
\text{where}\quad
\mu_{\mathrm{module}} &= \frac{1}{n}\sum_{j=1}^{n}\phi_j, \qquad
\sigma_{\mathrm{module}} = \sqrt{\frac{1}{n}\sum_{j=1}^{n}\left(\phi_j-\mu_{\mathrm{module}}\right)^2}.
\end{aligned}
\end{equation}

\noindent
where $\phi_i$ is the value of a specific feature (e.g., the capacity slope) for cell group $i$, and $\mu_{\mathrm{module}}$ and $\sigma_{\mathrm{module}}$ are the mean and standard deviation of that same feature across all $n=14$ cell groups within that specific module. The resulting metric $\phi^{\mathrm{norm}}_i$ represents the normalized feature value for the cell group.

The resulting normalized values place every cell group on a common scale, representing how many standard deviations its behavior deviates from its module's average. Figure~\ref{fig:normalization_example}(C) visually demonstrates how this normalization aligns the features from different modules, allowing us to effectively identify abnormal behaviors across the entire dataset.

\subsection{Statistical Separation of Faulty and Non-Faulty Groups with respect to Features}
\label{subsec:p-value_results}

\begin{figure}[h!]
    \centering
    \includegraphics[width=.95\linewidth]{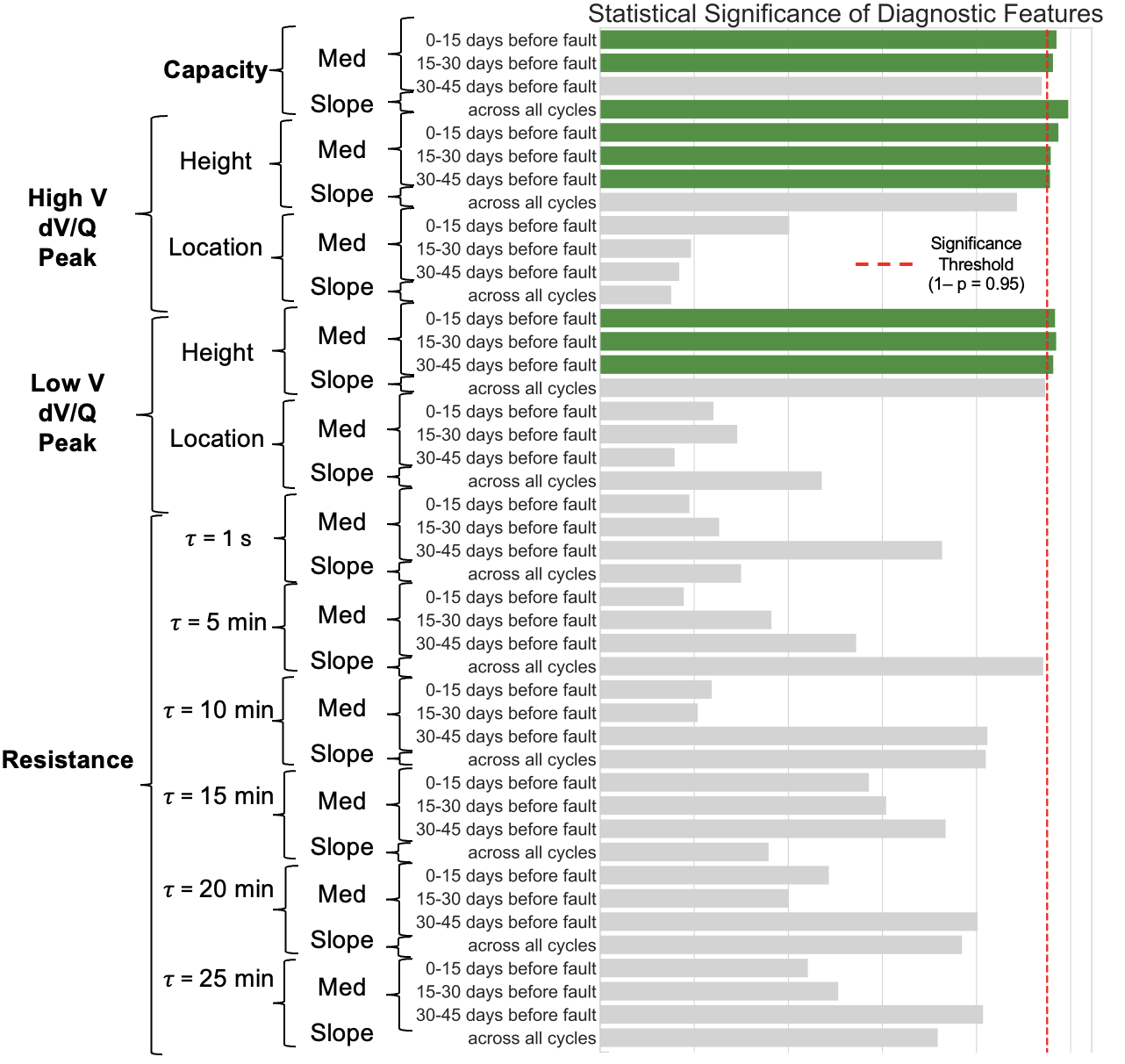}
    \caption[Statistical Significance of Normalized Health Features in Separating Faulty Groups]{Statistical Significance of Normalized Health Features in Separating Faulty Groups. Bar plots display $1 - p$-values from Mann-Whitney U test comparing normalized calibrated health features for faulty vs. non-faulty cell groups, with a significance threshold of $1 - p = 0.95$ indicated by the dashed line.}
    \label{fig:wong3}
\end{figure}

Figure~\ref{fig:wong3} summarizes the $p$-values for each feature from the Mann–Whitney U test (~\ref{subsec:mann_whitney}). A $p$-value $< 0.05$ indicates a statistically significant difference between the distributions of the faulty and non-faulty groups with at least 95\% confidence.

Among the DVA features, the peak heights for both the High V and Low V peaks differed significantly between faulty and non-faulty groups ($p < 0.05$), whereas peak locations and slopes did not ($p > 0.05$). Figure~\ref{fig:fault_result} (A)--(C) show that the median heights of both peaks were significantly lower for the faulty cell groups. Notably, this significance persisted across all the tested pre-fault time windows, with a clear separation emerging as early as 30-45 days prior to the fault.

Capacity-based features also demonstrated strong statistical separation. Both median capacity values and the capacity degradation rate showed a statistically significant difference between faulty and non-faulty cell groups ($p<0.05$). As shown in Figure~\ref{fig:fault_result} (D) and (E), faulty cell groups exhibited significantly lower median capacity as early as 15--30 days before the fault and a more negative degradation slope (indicating accelerated capacity loss).

In contrast to capacity and DVA features, internal resistance estimates showed no statistically significant difference between faulty and non-faulty groups ($p>0.05$). This result indicates that increased internal resistance was not a primary characteristic of specific faults captured in this dataset.

\subsection{Utility of Features for Fault Detection}
\label{subsec:auc_results}

We assessed the potential of individual health features for fault detection by calculating their Area Under the Receiver Operating Characteristic Curve (AUC). The ROC curve characterizes the trade-off between the true positive rate and the false positive rate as a decision threshold is swept across the feature values. An AUC of 0.5 indicates no discriminative power (equivalent to random guessing), while 1.0 indicates perfect separation between faulty and non-faulty populations (see ~\ref{sec:AUC_appendix}).

Despite the statistical significance reported in Section \ref{subsec:p-value_results}, the capacity and DVA features yielded AUC values that indicate substantial distributional overlap between the faulty and non-faulty cell groups. AUC values were consistently below 0.70, implying that these features alone cannot support robust fault detection without incurring a substantial false positive rate. The capacity degradation rate (slope) exhibited the highest discriminative capability (AUC = 0.671). Median capacity values were lower, ranging from 0.633 (0-15 days before fault) to 0.614 (30-45 days before fault). Similarly, the AUC for High V and Low V dV/dQ peak heights remained approximately 0.64 across all time windows.

The AUC values for resistance features were generally below 0.60. These scores approach the baseline of random guessing (0.5), confirming that resistance features lack the discriminative power necessary to effectively detect these faults.

\begin{figure} [h!]
    \centering
    \includegraphics[width=1\linewidth]{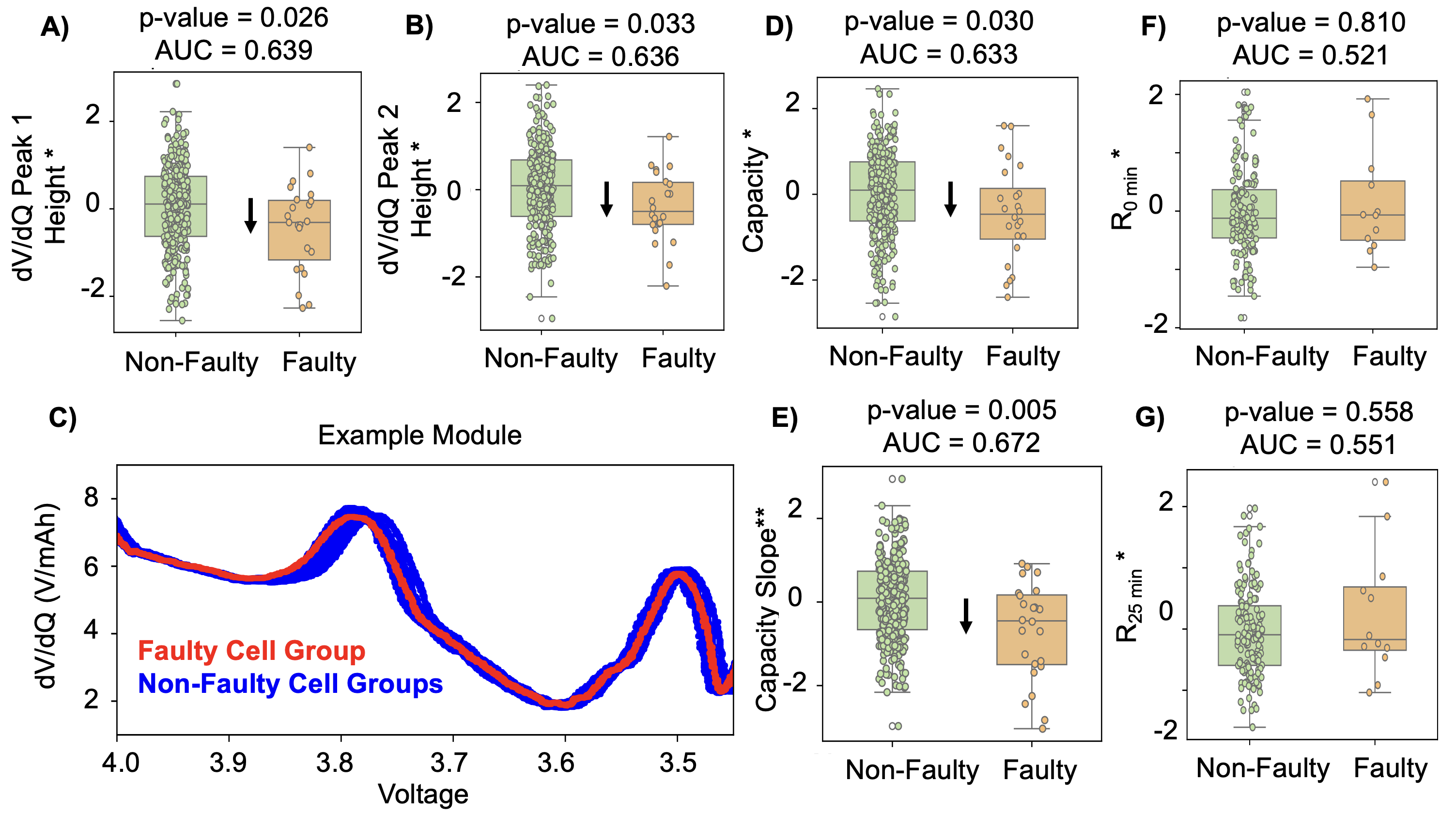}
    \caption[Comparison of Normalized Health Features for the Non-Faulty and Faulty Cell Group Populations]{Comparison of normalized health features for the non-faulty and faulty cell populations. Faulty cells show statistically significantly lower values for (A, B) dV/dQ peak heights, (D) capacity, and (E) capacity slope ($p < 0.05$). (C) dV/dQ curves of the cell groups within an exemplary module (module 1) 10 days before the fault. No statistical difference in (F) $R_{0 min}$, (G) $R_{25 min}$ ($p > 0.05$). $^*$ - median value over 0-15 days before the fault. $^{**}$ - the slope across all available cycles.}
    \label{fig:fault_result}
    \label{fig:fault_result}
\end{figure}

\subsection{Discussion}

\subsubsection{Hypothesized Failure Mechanisms Consistent with Fault Signatures}

The statistically significant differences in specific health features enables us to hypothesize potential physical mechanisms driving the faults.

The reduced dV/dQ peak heights observed in faulty cell groups are consistent with mechanisms such as cell-to-cell imbalances and inhomogeneities in the parallel group. The peaks of the dV/dQ curve correspond to phase transitions in the electrode materials;  peak broadening and the corresponding reduction in height signify greater imbalances and inhomogeneities~\cite{DUBARRY201919, Lewerenz2017, LEWERENZ2018421, Sieg2020}. Such imbalances and inhomogeneities can lead to uneven current distributions \cite{Fuhrmann_2024, WENG201336, FATH2019100813}, forcing some cells to operate at higher effective C-rates. This, in turn, can generate localized resistive heating that accelerates the degradation mechanism or, under severe conditions, trigger faults \cite{Marlow2024}. 

Furthermore, the faulty groups' lower capacity and accelerated degradation rate suggest that the faults could be associated with local lithium plating. Experimental studies by Yuan et al. \cite{Yuan2025}, Chen et al. \cite{Chen2024}, and Tian et al. \cite{Tian2023} have demonstrated that plating defects cause an irreversible loss of lithium inventory (LLI) that lead to batteries with local lithium plating defects to exhibit lower capacity and a steeper degradation trajectory compared to batteries with no plating. 

While post-mortem analysis would be required to confirm these mechanisms, the feature signatures observed here align with these established degradation modes.

A critical finding of this study is the inability of the resistance estimations to reliably distinguish faulty cell groups ($p>0.05$), which challenges conventional reliance on using resistance for fault detection. Previous studies document faults that are reflected in significant changes in resistances. For instance, connector loss and corrosion can lead to high contact resistance~\cite{Offer2012}. Internal short circuit faults and thermal faults can be characterized by ohmic internal resistance increases \cite{Feng2018a,Gao2019}.
However, the lack of statistical separation suggests that significant resistance deviation was not a primary characteristic of the specific faulty cell groups in our dataset. While this may indicate that the underlying fault mechanisms are not related to significant changes in resistance, it is also possible that the parallel connections mask the signal of a single cell with abnormal resistance. In a parallel configuration, the equivalent resistance is dominated by the low-resistance pathways of the healthy cells; consequently, a resistance increase in a single failing cell can be effectively hidden. This masking effect, specifically the low sensitivity of group-level resistance to individual cell degradation, has been quantified by researchers like Song et al.~\cite{Song2020}.

\subsubsection{Challenges in Fault Detection using only individual health features}
\label{subsubsec:limitations_future_work}
While our analysis establishes statistically significant differences in capacity and DVA-derived features between faulty and non-faulty groups, the distributional overlap highlights the challenges of utilizing these group-level features individually for robust fault detection in the field.  These limitations likely arise from two primary factors: the heterogeneity of the underlying fault mechanisms and the fundamental limits of observability of faults from electrochemical signals.

First, the root causes of the faults likely vary and consequently cause different electrochemical behaviors. As discussed in Section \ref{subsec:p-value_results}, a significant portion of the faulty groups exhibited reduced dV/dQ peak heights and lower capacity; these behaviors are consistent with mechanisms such as cell-to-cell imbalances, inhomogeneities, and lithium plating. However, the population of faulty groups likely included groups that failed via alternative mechanisms that do not exhibit these electrochemical behaviors~\cite{Hu2020}. Consequently, the performance of any individual feature for fault detection is diluted by the presence of faults that do not manifest the targeted electrochemical behaviors.

Second, some faults may be fundamentally unobservable from electrochemical signals. Recent studies indicate that even in an isolated single cell, localized mechanical or thermal failures, such as electrode fractures or early-stage internal short circuits, induce minimal cell voltage changes that are often indistinguishable from sensor noise~\cite{Fan2025}. These minimal voltage changes would be further attenuated if the single faulty cell is in parallel with healthy cells. Consequently, the fault signal would be unobservable through electrochemical signals regardless of the feature extraction method used.

\section{Conclusion}
\label{sec:conclusion}

Monitoring the health of lithium-ion battery modules from their operational data is critical for ensuring safe operation; however, this task is challenging due to the variability of operating conditions, which confounds health features, and the limited sensing in modules, which prevents the direct observation of individual cell states. This work developed and demonstrated a framework for extracting and calibrating health features for battery modules from available operational data. We applied this framework to field data from 25 commercial grid-connected lithium-ion BESS modules; each module consisted of 14 series-connected parallel groups, one of which was identified as faulty via post-mortem analysis. Using only the available current and voltage measurements, our framework first extracts capacity, resistance, and differential voltage analysis features for all cell groups within the modules. It then employs Gaussian process regression to calibrate the health features, separating features’ time-dependent trend indicative of degradation from the confounding effects of variable operating conditions.

With the calibrated health features, we evaluated the potential of these features to separate faulty parallel-connected cell groups within the modules. Our analysis demonstrated that calibrated capacity, capacity degradation rate, and dV/dQ peak heights separated the faulty cell groups with statistical significance ($p < 0.05$). These statistically significant differences suggest that the faults may be consistent with mechanisms such as cell-to-cell imbalances, inhomogeneities, and lithium plating. Conversely, internal resistance features ($\widehat{R}_{\mathrm{\mathrm{post\_chg\_rel}},\tau}$) did not show statistical separation ($p > 0.05$). This lack of separation was observed for $\tau \in \{1\,\text{s}, 5\,\text{min}, 10\,\text{min}, 15\,\text{min}, 20\,\text{min}, 25\,\text{min}\}$. These resistance estimates obtained at different values of $\tau$ capture distinct physical contributions: 
$\widehat{R}_{\mathrm{\mathrm{post\_chg\_rel}},\,1~\mathrm{s}}$ primarily reflects the ohmic resistance of the cell group, while resistance estimates obtained further into the relaxation period ($\tau > 1~\mathrm{s}$) increasingly incorporates charge-transfer and diffusion-related resistances. Our results indicate that the specific faults in this dataset were not distinguishable via any of these resistance components, challenging the conventional notion that faults are primarily reflected by significant changes in internal resistance. As discussed, resistance features were likely ineffective because the parallel architecture masks the signal of individual high-resistance cells. While the features highlighted statistically significant differences that provide insight into the potential physical mechanisms driving the faults, there were substantial distributional overlaps between faulty and non-faulty cell groups with respect to the features, highlighting the challenge of utilizing the features individually for robust fault detection in the field.

Ultimately, this work furthers the understanding of how cell health can be assessed from field data. By applying our framework to commercial BESS modules containing verified faulty parallel-connected cell groups, we demonstrate both the potential and the limitations of electrochemical signals for detecting faults in real-world operations.

% \section*{Author Contributions}
% your text here

\section*{CRediT authorship contribution statement}

\textbf{Clement Wong}: Conceptualization, Methodology, Software, Formal analysis, Investigation, Visualization, Writing – original draft. \textbf{Andrew Weng}: Writing – review \& editing. \textbf{Xin Hui Ooi}: Writing – review \& editing. \textbf{Zhiwen Wan}: Writing – review \& editing. \textbf{Jeesoon Choi}: Project administration.
\textbf{Seung Yoon Yang}: Project administration. \textbf{Heejun Jin}: Project administration. \textbf{Jason Siegel}:  Writing – review \& editing. \textbf{Anna Stefanopoulou}: Supervision, Writing – review \& editing, Funding acquisition.

\section*{Funding}
This work was supported by LG Energy Solution. 

% \section*{Declaration of Competing Interest}
% The authors declare that they have no known competing financial interests or personal relationships that could have appeared to influence the work reported in this paper.

% \vfill
% \break
\newpage

\appendix

\section{OCV–SOC curve of a fresh cell}
\label{subsec:fresh_OCV}

\begin{figure}[h!]
    \centering
    \includegraphics[width=0.7\linewidth]{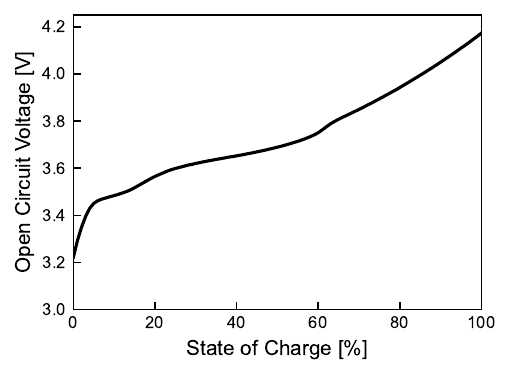}
    \caption[Laboratory-Measured OCV–SOC Curve of Fresh Cell]{Laboratory-measured OCV–SOC curve of fresh cell}
    \label{fig:OCV}
\end{figure}

\section {Gaussian Process (GP) Modeling Framework to isolate the change in a health feature vs time}
\label{sec:GP_modeling_framwork}

To isolate the change in a health feature vs time of each cell in a module from variations caused by operating conditions, we employ GP framework. As described in Section \ref{sec:feature_sensitivity_to_operation_condition}, a GP model is trained for each individual cell group within a module. Our approach allows the model to learn the effects of operating conditions on a health feature and effectively decouple these effects to uncover the underlying trend between a health feature and time indicative of degradation for a cell group in a module. 

\subsection{GP Model for a health feature}

For a given cell group, its health feature $y$ at time $t$ under operating conditions $x$ is modeled as the sum of a latent function $f$ and independent Gaussian noise $\epsilon$:

$$y(\mathbf{x}, t) = f(t, \mathbf{x}) + \epsilon, \quad \text{where} \quad \epsilon \sim \mathcal{N}(0, \sigma_n^2)$$

To isolate the degradation trend from operational effects, we structure $f(x(t),t)$ as the sum of two independent Gaussian Processes: one for the operating conditions ($f_{op}$) and one for time-dependent degradation ($f_{deg}$). 

\begin{equation} 
f(\mathbf{x}, t) = f_{op}(\mathbf{x}) + f_{deg}(t) 
\end{equation} 
\begin{equation} 
f_{op}(\mathbf{x}) \sim \mathcal{GP}(\mu_{op}(x), k_{op}(\mathbf{x}, \mathbf{x}')) \end{equation} 
\begin{equation} 
f_{deg}(t) \sim \mathcal{GP}(\mu_{deg}(t), k_{deg}(t, t')) 
\end{equation}

Because the two component GPs are independent, the mean and kernel of the combined process are simply the sum of their parts.
\begin{equation}
    \mu(t,x) = \mu_{op}(x) + \mu_{deg}(t)
\end{equation}
\begin{equation}
k((x, t), (x', t')) = k_{op}(\mathbf{x}, \mathbf{x}') + k_{deg}(t, t')
\end{equation}

% The latent function $f$ is modeled as a GP, $f \sim \mathcal{GP}(\mu, k)$,
% with a kernel structured as a sum of two kernels: one to learn the sensitivity of the health feature to operating conditions  ($k_{op}$) and one to the cell-specific degradation trend described by the health feature ($k_{deg}$).

% \begin{equation}
% k((x, t), (x', t'')) = k_{op}(\mathbf{x}, \mathbf{x}') + k_{deg}((t), (t',c'))
% \end{equation}

\subsubsection{Operational Variability Kernel ($k_{op}$)}
For $k_{op}(\mathbf{x}, \mathbf{x}')$, we use a radial basis function (RBF) kernel in modeling the dependency of a health feature on operating conditions, assuming that the variability of a health feature to operating conditions are relatively smooth.
\begin{equation}
k_{op}(\mathbf{x}, \mathbf{x}') = \sigma_{op}^2 \exp\left(-\frac{1}{2} \sum_{d=1}^{2} \frac{(x - x')^2}{\ell^2}\right) 
\end{equation}

\subsubsection{Cell-Specific Degradation Kernel ($k_{deg}$)}

To model a unique degradation rate for each cell, we define a kernel $k_{deg}( t, t' )$ using a non-stationary Brownian motion kernel. This kernel is selected because its cumulative properties provide a direct mathematical analogue for the irreversible, accumulating process of battery degradation, as discussed in Section \ref{sec:GP_modeling_framwork}.
\begin{equation}
k_{deg}( t, t' ) = \min(t, t')
\end{equation}

\subsection{Model Implementation}

The model was trained using the GPy library in Python. The training process for each cell group's model involved two key steps: data normalization and hyperparameter optimization.

\subsubsection{Data normalization}
Prior to training each model, its specific inputs (operating conditions) and output (health feature $y$) were normalized to have zero mean and unit variance. This standard pre-processing step ensures that the kernel computations are not dominated by any single input's scale and improves the stability of the numerical optimization.

\subsubsection{Hyperparameter optimization} 
The GP model's kernels are governed by a set of hyperparameters (e.g., lengthscales and variances) that were learned from the data for each health feature model. Using the GPy library, we optimized these hyperparameters by maximizing the marginal log-likelihood of the observations for that specific cell group, a procedure that finds the kernel parameters that best explain the data while inherently penalizing model complexity. Furthermore, we incorporated prior physical knowledge by setting scientifically reasonable bounds on the lengthscale hyperparameters before optimization, which constrained the model to learn relationships consistent with established battery behavior. The optimization was performed using the L-BFGS-B algorithm.

\section{Derivation of Resistance Sensitivity to Current} \label{appendix:resistance_derivation}

The observed decrease in resistance with increasing current magnitude can be explained by electrochemical processes governed by Butler-Volmer kinetics. The relationship between current density ($j$) and activation overpotential ($\eta$) is described by the Butler-Volmer equation:

\begin{equation}
    j = j_0 \left[ \exp\left( \frac{\alpha_a n F \eta}{RT} \right) - \exp\left( - \frac{\alpha_c n F \eta}{RT} \right) \right]
    \label{eq:butler_volmer}
\end{equation}

\noindent where $j_0$ is the exchange current density, $\alpha_a$ and $\alpha_c$ are the anodic and cathodic transfer coefficients, $n$ is the number of electrons involved in the reaction, $F$ is Faraday's constant, $R$ is the ideal gas constant, and $T$ is the temperature.

Charge transfer resistance ($R_{ct}$) is defined as:

\begin{equation}
    R_{ct} = \frac{d\eta}{dj}.
    \label{eq:r_def}
\end{equation}

For the resistance measurements performed in this study, the current pulses exceeded 0.1 C (Section \ref{subsec:implementation_details_resistance}). Under these high-current conditions, the overpotential $\eta$ becomes large. Consequently, the cathodic reaction term in Equation \ref{eq:butler_volmer} becomes negligible. The kinetics can thus be approximated by the Tafel equation:

\begin{equation}
    j \approx j_0 \exp\left( \frac{\alpha_a n F \eta}{RT} \right).
    \label{eq:tafel}
\end{equation}

Rearranging Equation (\ref{eq:tafel}) expresses the overpotential $\eta$ as a function of current:

\begin{equation}
    \eta = \frac{RT}{\alpha_a n F} \ln\left( \frac{j}{j_0} \right).
    \label{eq:tafel_rearranged}
\end{equation}

Differentiating $\eta$ with respect to $j$ in Equation (\ref{eq:tafel_rearranged}) yields $R_{ct}$ according to Equation (\ref{eq:r_def}):

\begin{equation}
    R_{ct} = \frac{d}{dj} \left( \frac{RT}{\alpha_a n F} \ln\left( \frac{j}{j_0} \right) \right) = \frac{RT}{\alpha_a n F} \cdot \frac{1}{j}.
    \label{eq:Rct_current}
\end{equation}

As shown in Equation (\ref{eq:Rct_current}), the charge transfer resistance is inversely proportional to the current magnitude ($R_{ct} \propto j^{-1}$), explaining the monotonic decrease in resistance observed in the field data.

\section{Derivation of Capacity Estimation Sensitivity to ${V}_{\mathrm{post\_chg\_rel},\,\mathrm{end}}$}
\label{subsec:supporting_work_for_cap_vs_vstart}

As presented in Section \ref{subsec:cap_estimation_method}, the capacity estimate is defined as:

\begin{equation}
    \widehat{C} = \frac{Q_{\mathrm{dis}}}{\widehat{z}_{\mathrm{dis},\,\mathrm{start}} - \widehat{z}_{\mathrm{dis},\,\mathrm{end}}}
\end{equation}

\noindent where $\widehat{z}_{\mathrm{dis},\,\mathrm{start}} = \mathrm{OCV}_{\mathrm{fresh}}^{-1}({V}_{\mathrm{post\_chg\_rel},\,\mathrm{end}})$ and $\widehat{z}_{\mathrm{dis},\,\mathrm{end}} = \mathrm{OCV}_{\mathrm{fresh}}^{-1}({V}_{\mathrm{post\_dis\_rel},\,\mathrm{end}})$.

${V}_{\mathrm{post\_chg\_rel},\,\mathrm{end}}$ is subject to measurement error $\epsilon$ relative to the true  open-circuit voltage at the start of discharge.
\begin{equation}
{V}_{\mathrm{post\_chg\_rel},\,\mathrm{end}} = {\mathrm{V}}_{\mathrm{\mathrm{post\_chg\_rel}},\,\mathrm{end}, \mathrm{true}} + \epsilon.
\end{equation}

We approximate $\widehat{z}_{\mathrm{dis},\,\mathrm{start}}$ using a first-order Taylor series expansion around $\mathrm{V}_{\mathrm{\mathrm{post\_chg\_rel}},\,\mathrm{end}, \mathrm{ true}} $:

\begin{equation}
    \widehat{z}_{\mathrm{dis},\,\mathrm{start}} = \mathrm{OCV}_{\mathrm{fresh}}^{-1}(\mathrm{V}_{\mathrm{\mathrm{post\_chg\_rel}},\,\mathrm{end,}\,\mathrm{true}} + \epsilon) \approx z_{\mathrm{dis},\,\mathrm{start,}\,\mathrm{true}} + \frac{d(\mathrm{OCV}_{\mathrm{fresh}}^{-1})}{dV} \cdot \epsilon
\end{equation}

Applying the Inverse Function Theorem, the derivative of the inverse OCV function is the reciprocal of the OCV slope, denoted here as $S(V) = \frac{d(\mathrm{OCV}_{\mathrm{fresh}})}{dz}$ . Substituting this into the expansion yields:

\begin{equation}
    \widehat{z}_{\mathrm{dis},\,\mathrm{start}} \approx z_{\mathrm{dis,start, true}} + \frac{\epsilon}{S({V}_{\mathrm{post\_chg\_rel},\,\mathrm{end}})}
\end{equation}

A similar expansion applies to $\widehat{z}_{\mathrm{dis},\,\mathrm{end}}$:

\begin{equation}
    \widehat{z}_{\mathrm{dis},\,\mathrm{end}} \approx z_{\mathrm{dis,end, true}} + \frac{\epsilon}{S({V}_{\mathrm{post\_dis\_rel},\,\mathrm{end}})}
\end{equation}

\begin{figure}[h!]
    \centering
    \includegraphics[width=.8\linewidth]{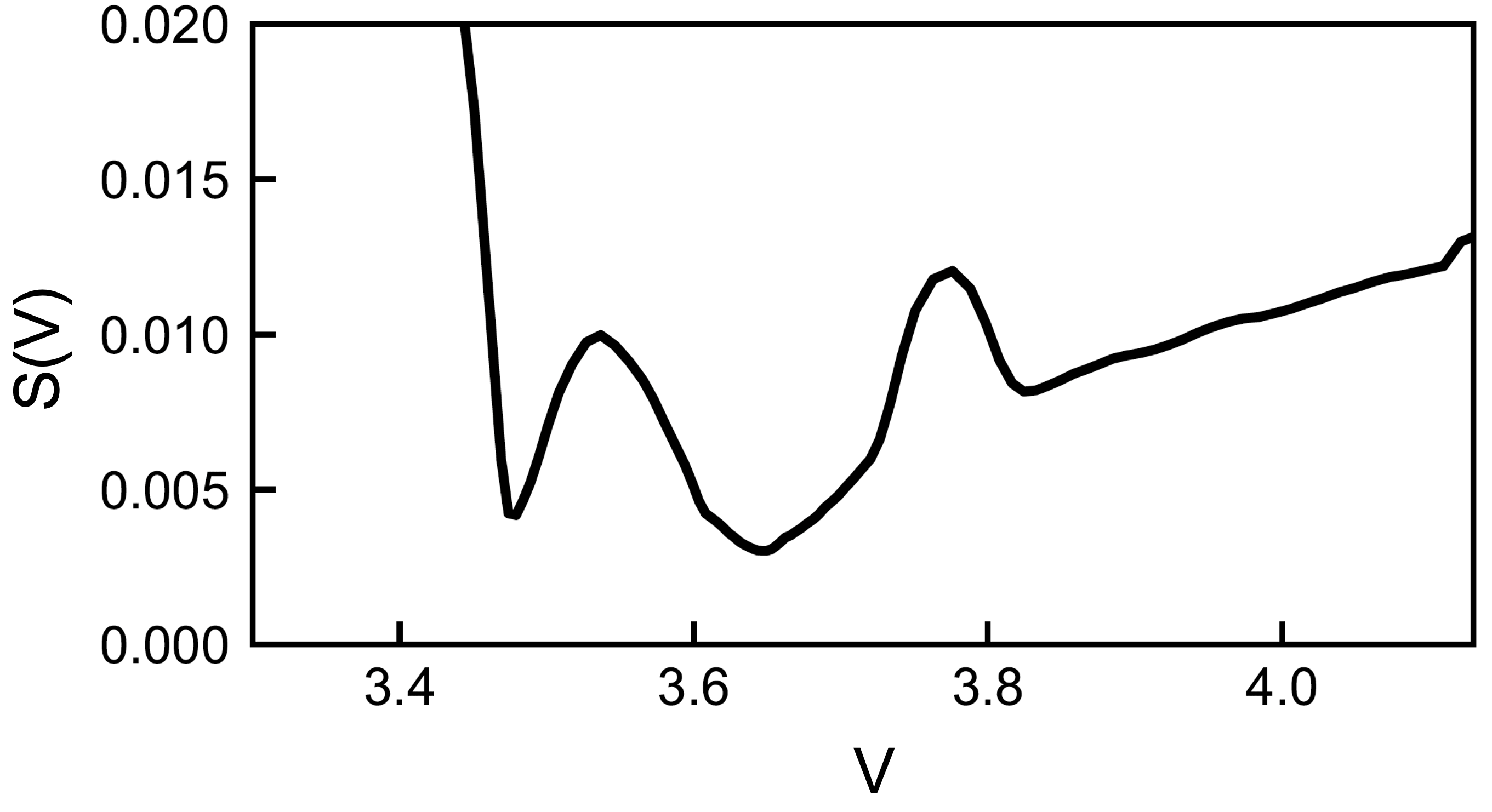}
    \caption[Derivative of the Fresh Cell OCV Curve with respect to SOC vs Voltage]{$S(V)$ vs V where $S(V) = \frac{d(\mathrm{OCV}_{\mathrm{fresh}})}{dz}$}
    \label{fig:dOCV}
\end{figure}

Therefore, the change in SOC is:

\begin{equation}
\label{equation_27}
\begin{split}
    \widehat{z}_{\mathrm{dis},\,\mathrm{start}} - \widehat{z}_{\mathrm{dis},\,\mathrm{end}} &= \left(z_{\mathrm{dis,start, true}} + \frac{\epsilon}{S({V}_{\mathrm{post\_chg\_rel},\,\mathrm{end}})}\right) \\&- \left(z_{\mathrm{dis,end, true}} + \frac{\epsilon}{S({V}_{\mathrm{post\_dis\_rel},\,\mathrm{end}})}\right) \\
    &= (z_{\mathrm{dis,start, true}} - z_{\mathrm{dis,end, true}}) \\&+ \epsilon \left(\frac{1}{S({V}_{\mathrm{post\_chg\_rel},\,\mathrm{end}})}-\frac{1}{S({V}_{\mathrm{post\_dis\_rel},\,\mathrm{end}})}\right)
\end{split}
\end{equation}
Figure \ref{fig:dOCV} shows $S(V)$ across voltages.

${V}_{\mathrm{post\_dis\_rel},\,\mathrm{end}}$ is typically located in the lower knee region of the OCV curve, where the OCV gradient is steep and $S({V}_{\mathrm{post\_dis\_rel},\,\mathrm{end}}) \gg S({V}_{\mathrm{post\_chg\_rel},\,\mathrm{end}})$. Consequently, the second term in the parenthesis becomes negligible, and the equation simplifies to:

\begin{equation}
\label{eq:SOC_diff}
    \widehat{z}_{\mathrm{dis},\,\mathrm{start}} - \widehat{z}_{\mathrm{dis},\,\mathrm{end}} 
    \approx (z_{\mathrm{dis,start, true}} - z_{\mathrm{dis,end, true}}) + \frac{\epsilon}{S({V}_{\mathrm{post\_chg\_rel},\,\mathrm{end}})}
\end{equation}

Substituting Equation \ref{eq:SOC_diff} into the capacity equation, we arrive at:

\begin{equation}
\label{eq:capacity_with_measurement_error}
    \widehat{C} \approx \frac{Q_{\mathrm{dis}}}{(z_{\mathrm{dis,start, true}} - z_{\mathrm{dis,end, true}}) + \frac{\epsilon}{S({V}_{\mathrm{post\_chg\_rel},\,\mathrm{end}})}}
\end{equation}

Equation \ref{eq:capacity_with_measurement_error} demonstrates that the capacity estimate is dependent on the local OCV gradient at the start of discharge. A value of ${V}_{\mathrm{post\_chg\_rel},\,\mathrm{end}}$ located in a steeper OCV region leads to a higher capacity estimate:

\begin{equation*}
    S({V}_{\mathrm{post\_chg\_rel},\,\mathrm{end}}) \uparrow \implies \frac{\epsilon}{S({V}_{\mathrm{post\_chg\_rel},\,\mathrm{end}})}  \downarrow \implies \widehat{C} \uparrow
\end{equation*}

\section{Intra-Module Uniformity of Operating Conditions}
\label{subsec:Intra-Module Uniformity}

\begin{figure}[h!]
    \centering
    \includegraphics[width=0.8\linewidth]{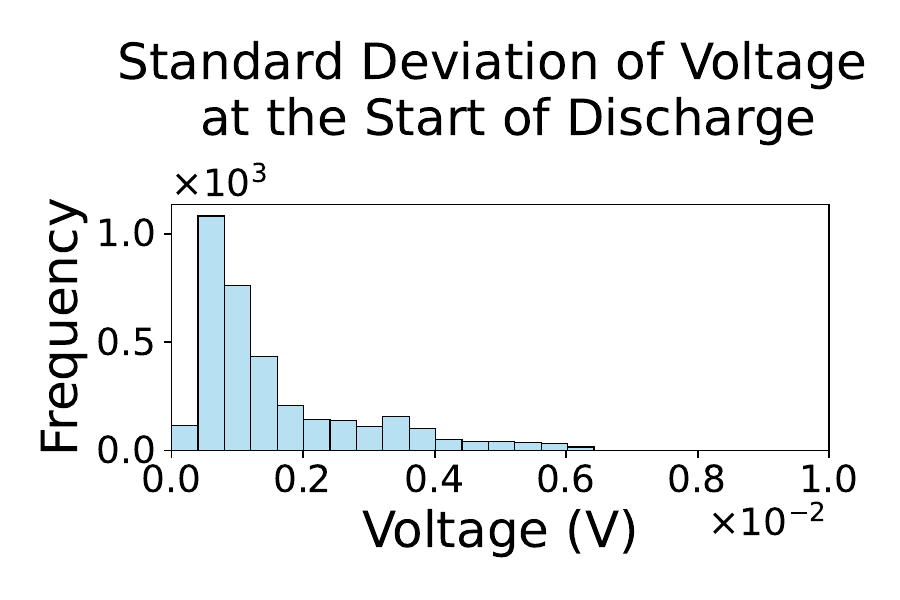}
    \caption[Distribution of Intra-Module Voltage Standard Deviation]{Distribution of Intra-Module Voltage Standard Deviation. The histogram shows the standard deviation of the 14 cell group voltages at the end of the post-charge relaxation period, aggregated across all cycles and all 25 modules. The sharp peak at low values indicates that cell groups within each module are consistently well-balanced in voltage after charging.}
    \label{fig:std_voltage_module}
\end{figure}

Within each module, operating conditions are more consistent across cell groups, which helps with comparing the health features of cell groups. The battery management system actively balances the cell groups after charging, resulting in minimal deviation in voltage across cell groups, as shown in Figure~\ref{fig:std_voltage_module}. Additionally, because the cell groups are connected in series, they share the same current. This consistency in both voltage after charge and current ensures that any significant deviation in a health feature from one cell group relative to others in the same module likely reflects a change in intrinsic health rather than operational differences.

\section{Mann Whitney U test}
\label{subsec:mann_whitney}
A Mann–Whitney U test for two samples was employed in this study to assess whether differences in the features between fault cells and non-fault cells were statistically significant. This test is particularly suitable when the assumptions of normality may not hold, or when there are substantial differences in sample sizes. The p-value obtained from the Mann–Whitney U test quantifies the level of marginal significance within the statistical hypothesis test, representing the probability that the null hypothesis—namely, that there is no difference in the distributions of the two groups—holds true. A p-value less than 0.05 was used to reject this null hypothesis, indicating a statistically significant difference. Because the Mann–Whitney U test does not rely on the data being normally distributed, it offers robust results even for skewed data and/or presence of outliers. In addition, box-and-whisker plots were used throughout the study to visually summarize the distributions of measured outcomes

\section{Methodology for Evaluating Fault Classification Performance of Individual Features}
\label{sec:AUC_appendix}

\subsection{Threshold-Based Decision Rule}

For a given health feature value $x$ and a specific decision threshold $\tau$, we assign the predicted class label ($\hat{y}$) using the following logic:

\begin{equation}
    \hat{y} = 
    \begin{cases} 
      \text{Faulty} & \text{if } x < \tau \\
      \text{Non-Faulty} & \text{if } x \geq \tau 
    \end{cases}
\end{equation}

For features where faults are characterized by an increase in value (e.g., Internal Resistance), the inequality direction is reversed ($x > \tau$ implies Faulty).

\begin{figure}[h]
    \centering
    \includegraphics[width=0.5\linewidth]{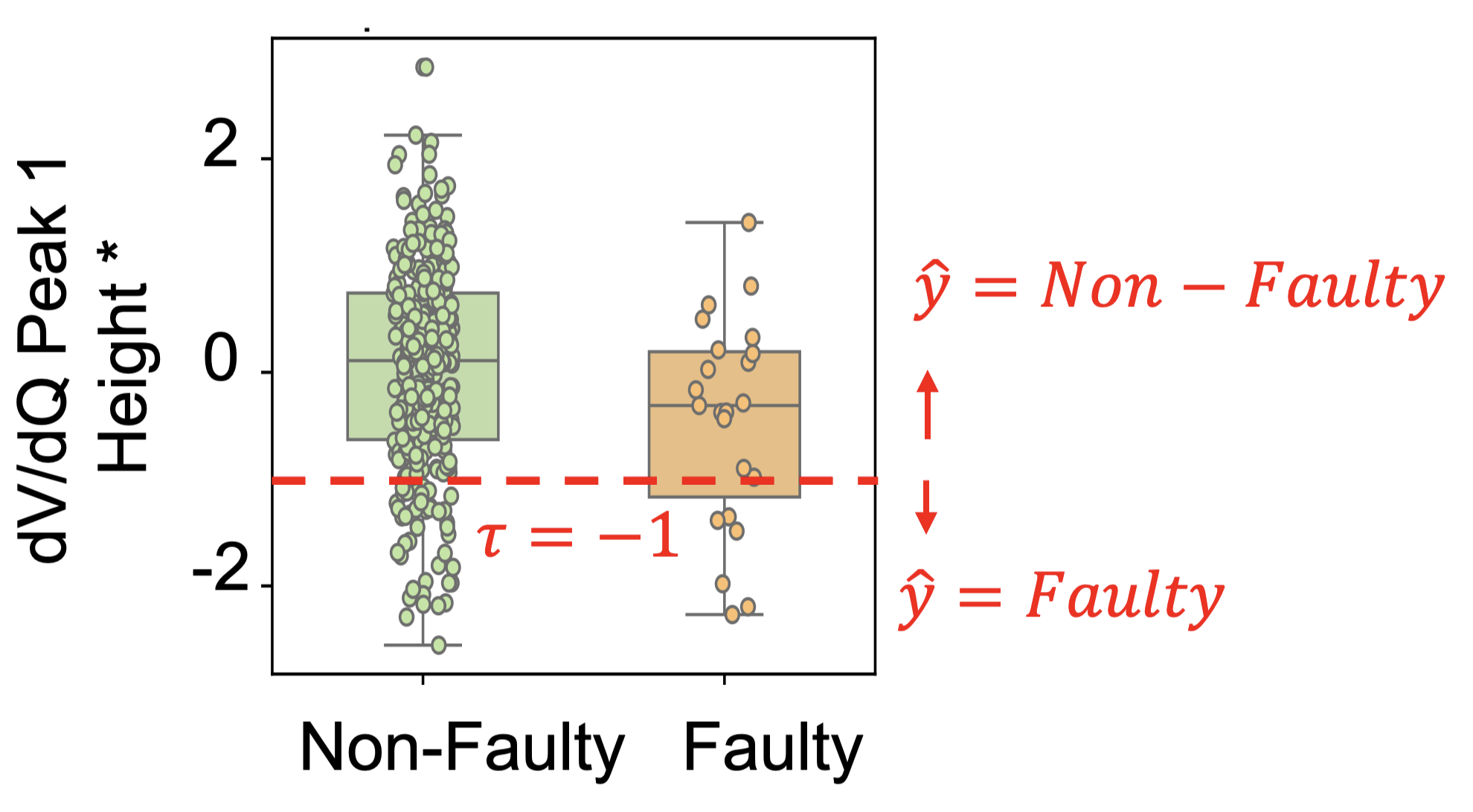}
    \caption[Threshold-Based Decision Rule for ROC Curve Generation]{Threshold-based decision rule applied to example feature (i.e. $x =$  dV/dQ Peak 1 Height $^*$) where $\tau$ = -1. $^*$ - median value over 0-15 days before the fault}
    \label{fig:decision-based-rule}
\end{figure}

By comparing these predicted labels ($\hat{y}$) against the true labels for every cell group in the dataset, we calculate the True Positive Rate (TPR) and False Positive Rate (FPR) for a specific threshold $\tau$:
\begin{equation}
    TPR(\tau) = \frac{\text{True Positives }(\tau)}{\text{True Positives }(\tau) + \text{False Negatives }(\tau)}
\end{equation}
\begin{equation}
    FPR(\tau) = \frac{\text{False Positives }(\tau)}{\text{False Positives }(\tau) + \text{True Negatives }(\tau)}
\end{equation}

\subsection{Receiver Operating Characteristic (ROC) and Area Under the Curve (AUC)}
The Receiver Operating Characteristic (ROC) curve plots $TPR(\tau)$ against $FPR(\tau)$ with the threshold $\tau$ ranging from $-\infty$ to $+\infty$. This curve visualizes the trade-off between detection capability and false alarm rates for every possible sensitivity setting.

The AUC is the integral of the ROC curve:

\begin{equation}
    AUC = \int_{0}^{1} TPR(FPR) \, d(FPR)
\end{equation}

\begin{figure}
    \centering
    \includegraphics[width=0.5\linewidth]{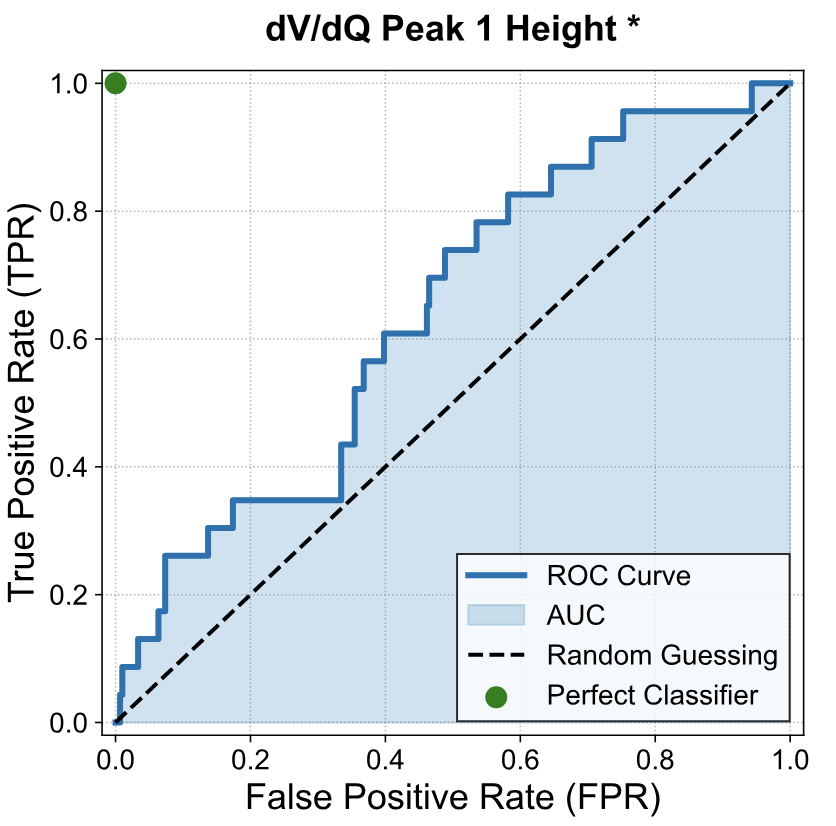}
    \caption[ROC and AUC Analysis for an Example Health Feature]{ROC and AUC analysis for an example feature (i.e. dV/dQ Peak 1 Height $^*$. $^*$ - median value over 0-15 days before the fault) }
    \label{fig:ROC_AUC}
\end{figure}

The AUC represents the probability that a randomly selected faulty cell group will have a more abnormal feature value than a randomly selected non-faulty cell group. The values of AUC, which ranges from 0 to 1, ranges evaluates the classification performance:  
\begin{itemize}
    \item AUC = 0.5: The feature has no discriminative power; its performance is equivalent to random guessing.
    \item 0.5 $<$ AUC $<$ 1.0: The feature provides better-than-random classification. A higher value indicates less overlap between the distributions of the faulty and non-faulty populations.
    \item AUC = 1.0: Perfect separation. There exists a threshold $\tau$ that perfectly separates faulty and non-faulty groups with 100\% sensitivity and 0\% false positives.
\end{itemize}

\newpage

\bibliographystyle{elsarticle-harv} 
\bibliography{main}

@book{GP_textbook,
author = {Rasmussen, Carl Edward and Williams, Christopher K I},
doi = {10.7551/mitpress/3206.001.0001},
isbn = {9780262256834},
publisher = {The MIT Press},
title = {{Gaussian Processes for Machine Learning}},
url = {https://doi.org/10.7551/mitpress/3206.001.0001},
year = {2005}
}

@article{Chen2024,
author = {Chen, Yingjie and Zhang, Huaqin and Hong, Jichao and Hou, Yankai and Yang, Jingsong and Zhang, Chi and Ma, Shikun and Zhang, Xinyang and Yang, Haixu and Liang, Fengwei and Li, Kerui},
doi = {10.1016/j.energy.2024.131574},
issn = {18736785},
journal = {Energy},
number = {March},
pages = {131574},
publisher = {Elsevier Ltd},
title = {{Lithium plating detection of lithium-ion batteries based on the improved variance entropy algorithm}},
url = {https://doi.org/10.1016/j.energy.2024.131574},
volume = {299},
year = {2024}
}

@article{Tian2023,
author = {Tian, Yu and Lin, Cheng and Chen, Xiang and Yu, Xiao and Xiong, Rui and Zhang, Qiang},
doi = {10.1016/j.ensm.2023.01.035},
issn = {24058297},
journal = {Energy Storage Materials},
number = {December 2022},
pages = {412--423},
publisher = {Elsevier B.V.},
title = {{Reversible lithium plating on working anodes enhances fast charging capability in low-temperature lithium-ion batteries}},
url = {https://doi.org/10.1016/j.ensm.2023.01.035},
volume = {56},
year = {2023}
}

@INPROCEEDINGS{Mohtat_2017,
  author={Mohtat, Peyman and Nezampasandarbabi, Farinaz and Mohan, Shankar and Siegel, Jason B. and Stefanopoulou, Anna G.},
  booktitle={2017 American Control Conference (ACC)}, 
  title={On identifying the aging mechanisms in li-ion batteries using two points measurements}, 
  year={2017},
  volume={},
  number={},
  pages={98-103},
  doi={10.23919/ACC.2017.7962937}}

@article{Zhou2025,
author = {Zhou, Zihao and Aitio, Antti and Howey, David},
doi = {10.1016/j.apenergy.2025.126375},
issn = {0306-2619},
journal = {Applied Energy},
number = {July},
pages = {126375},
publisher = {Elsevier Ltd},
title = {{Learning Li-ion battery health and degradation modes from data with aging-aware circuit models}},
url = {https://doi.org/10.1016/j.apenergy.2025.126375},
volume = {397},
year = {2025}
}

@article{Koch2018,
author = {Koch, Sascha and Birke, Kai Peter and Kuhn, Robert},
doi = {10.3390/batteries4020016},
pages = {1--11},
title = {{Fast Thermal Runaway Detection for Lithium-Ion Cells in Large Scale Traction Batteries}},
year = {2018}
}

@article{Fan2025,
author = {Fan, Jinbao and Liu, Chenchen and Li, Na and Yang, Le and Yang, Xiao-Guang and Dou, Bowen and Hou, Shujuan and Feng, Xuning and Jiang, Hanqing and Li, Hong and Song, Wei-Li and Sun, Lei and Chen, Hao-Sen and Gao, Huajian and Fang, Daining},
doi = {10.1038/s41586-025-08785-7},
issn = {1476-4687},
journal = {Nature},
number = {8063},
pages = {639--645},
title = {{Wireless transmission of internal hazard signals in Li-ion batteries}},
url = {https://doi.org/10.1038/s41586-025-08785-7},
volume = {641},
year = {2025}
}

@article{Bai2014,
author = {Bai, Peng and Bazant, Martin Z},
doi = {10.1038/ncomms4585},
issn = {2041-1723},
journal = {Nature Communications},
number = {1},
pages = {3585},
title = {{Charge transfer kinetics at the solid–solid interface in porous electrodes}},
url = {https://doi.org/10.1038/ncomms4585},
volume = {5},
year = {2014}
}

@INPROCEEDINGS{Ansean2013,
  author={Anseán, D. and García, V.M. and González, M. and Viera, J.C. and Blanco, C. and Antuña, J.L.},
  booktitle={2013 World Electric Vehicle Symposium and Exhibition (EVS27)}, 
  title={DC internal resistance during charge: Analysis and study on LiFePO4 batteries}, 
  year={2013},
  volume={},
  number={},
  pages={1-11},
  doi={10.1109/EVS.2013.6914746}}

@article{Luder2025,
author = {Luder, Daniel and John, Praise Thomas and Busch, Paul and B{\"{o}}rner, Martin and Cao, Wenjiong and Dechent, Philipp and Barbers, Elias and Bihn, Stephan and Liu, Lishuo and Feng, Xuning and Sauer, Dirk Uwe and Li, Weihan},
doi = {10.1016/j.geits.2025.100282},
issn = {27731537},
journal = {Green Energy and Intelligent Transportation},
number = {3},
title = {{Big data generation platform for battery faults under real-world variances}},
volume = {4},
year = {2025}
}

@article{Yi2013,
author = {Yi, Jaeshin and Kim, Ui Seong and Shin, Chee Burm and Han, Taeyoung and Park, Seongyong},
doi = {10.1149/2.039303jes},
issn = {0013-4651},
journal = {Journal of The Electrochemical Society},
number = {3},
pages = {A437--A443},
title = {{Three-Dimensional Thermal Modeling of a Lithium-Ion Battery Considering the Combined Effects of the Electrical and Thermal Contact Resistances between Current Collecting Tab and Lead Wire}},
volume = {160},
year = {2013}
}

@article{Han2020,
author = {Han, Seungyun and Choi, Changki and Kwon, Sanguk and Lee, Seongjun and Kim, Jonghoon},
doi = {10.23919/ICCAS50221.2020.9268389},
isbn = {9788993215205},
issn = {15987833},
journal = {International Conference on Control, Automation and Systems},
number = {Iccas},
pages = {496--498},
title = {{Electrical analysis about internal short circuit using additional resistance in high energy lithium-ion battery}},
volume = {2020-Octob},
year = {2020}
}

@article{Chen2019a,
author = {Chen, Mingbiao and Bai, Fanfei and Lin, Shili and Song, Wenji and Li, Yang and Feng, Ziping},
doi = {10.1016/j.applthermaleng.2018.10.011},
issn = {13594311},
journal = {Applied Thermal Engineering},
number = {October 2018},
pages = {775--784},
publisher = {Elsevier},
title = {{Performance and safety protection of internal short circuit in lithium-ion battery based on a multilayer electro-thermal coupling model}},
url = {https://doi.org/10.1016/j.applthermaleng.2018.10.011},
volume = {146},
year = {2019}
}

@article{Im2023,
author = {Im, Dong Hyeon and Chung, Ji Bum},
doi = {10.1016/j.est.2023.108192},
issn = {2352152X},
journal = {Journal of Energy Storage},
number = {February},
pages = {108192},
publisher = {Elsevier Ltd},
title = {{Social construction of fire accidents in battery energy storage systems in Korea}},
url = {https://doi.org/10.1016/j.est.2023.108192},
volume = {71},
year = {2023}
}

@article{Ward2022,
archivePrefix = {arXiv},
arxivId = {2109.07278},
author = {Ward, Logan and Babinec, Susan and Dufek, Eric J. and Howey, David A. and Viswanathan, Venkatasubramanian and Aykol, Muratahan and Beck, David A.C. and Blaiszik, Benjamin and Chen, Bor Rong and Crabtree, George and Clark, Simon and {De Angelis}, Valerio and Dechent, Philipp and Dubarry, Matthieu and Eggleton, Erica E. and Finegan, Donal P. and Foster, Ian and Gopal, Chirranjeevi Balaji and Herring, Patrick K. and Hu, Victor W. and Paulson, Noah H. and Preger, Yuliya and Uwe-Sauer, Dirk and Smith, Kandler and Snyder, Seth W. and Sripad, Shashank and Tanim, Tanvir R. and Teo, Linnette},
doi = {10.1016/j.joule.2022.08.008},
eprint = {2109.07278},
issn = {25424351},
journal = {Joule},
number = {10},
pages = {2253--2271},
publisher = {Elsevier Inc.},
title = {{Principles of the Battery Data Genome}},
url = {https://doi.org/10.1016/j.joule.2022.08.008},
volume = {6},
year = {2022}
}

@article{Attia2022,
author = {Attia, Peter M. and Bills, Alexander and {Brosa Planella}, Ferran and Dechent, Philipp and dos Reis, Gon{\c{c}}alo and Dubarry, Matthieu and Gasper, Paul and Gilchrist, Richard and Greenbank, Samuel and Howey, David and Liu, Ouyang and Khoo, Edwin and Preger, Yuliya and Soni, Abhishek and Sripad, Shashank and Stefanopoulou, Anna G. and Sulzer, Valentin},
doi = {10.1149/1945-7111/ac6d13},
issn = {0013-4651},
journal = {Journal of The Electrochemical Society},
number = {6},
pages = {060517},
publisher = {IOP Publishing},
title = {{Review—“Knees” in Lithium-Ion Battery Aging Trajectories}},
volume = {169},
year = {2022}
}

@article{Berecibar2016b,
author = {Berecibar, M. and Gandiaga, I. and Villarreal, I. and Omar, N. and {Van Mierlo}, J. and {Van Den Bossche}, P.},
doi = {10.1016/j.rser.2015.11.042},
issn = {18790690},
journal = {Renewable and Sustainable Energy Reviews},
pages = {572--587},
publisher = {Elsevier Ltd},
title = {{Critical review of state of health estimation methods of Li-ion batteries for real applications}},
volume = {56},
year = {2016}
}

@article{Xiong2018,
author = {Xiong, Rui and Li, Linlin and Tian, Jinpeng},
doi = {10.1016/j.jpowsour.2018.10.019},
issn = {03787753},
journal = {Journal of Power Sources},
number = {5},
pages = {18--29},
publisher = {Elsevier},
title = {{Towards a smarter battery management system: A critical review on battery state of health monitoring methods}},
url = {https://doi.org/10.1016/j.jpowsour.2018.10.019},
volume = {405},
year = {2018}
}

@article{Wong_2024,
author = {Wong, Clement and Weng, Andrew and Movahedi, Hamidreza and Choi, Jeesoon and Yang, Seung Yoon and Jin, Heejun and Siegel, Jason and Stefanopoulou, Anna},
doi = {https://doi.org/10.1016/j.est.2026.120507},
issn = {2352-152X},
journal = {Journal of Energy Storage},
pages = {120507},
title = {{Quantifying imbalances in parallel-connected cell groups using group voltage and current}},
url = {https://www.sciencedirect.com/science/article/pii/S2352152X26001714},
volume = {150},
year = {2026}
}

@article{Schaeffer2024,
archivePrefix = {arXiv},
arxivId = {2406.19015},
author = {Schaeffer, Joachim and Lenz, Eric and Gulla, Duncan and Bazant, Martin Z. and Braatz, Richard D. and Findeisen, Rolf},
doi = {10.1016/j.xcrp.2024.102258},
eprint = {2406.19015},
issn = {26663864},
journal = {Cell Reports Physical Science},
number = {11},
pages = {102258},
publisher = {The Author(s)},
title = {{Gaussian-process-based online health monitoring and fault analysis of lithium-ion battery systems from field data}},
url = {https://doi.org/10.1016/j.xcrp.2024.102258},
volume = {5},
year = {2024}
}

@article{Fuhrmann_2024,
author = {Fuhrmann, Marion and Torcheux, Laurent and Kobayashi, Yo},
doi = {https://doi.org/10.1016/j.jpowsour.2024.235210},
issn = {0378-7753},
journal = {Journal of Power Sources},
pages = {235210},
title = {{Knee point prediction for lithium-ion batteries using differential voltage analysis and degree of inhomogeneity}},
url = {https://www.sciencedirect.com/science/article/pii/S0378775324011625},
volume = {621},
year = {2024}
}

@article{Lewerenz2017,
author = {Lewerenz, Meinert and Marongiu, Andrea and Warnecke, Alexander and Sauer, Dirk Uwe},
doi = {10.1016/j.jpowsour.2017.09.059},
issn = {03787753},
journal = {Journal of Power Sources},
pages = {57--67},
publisher = {Elsevier B.V},
title = {{Differential voltage analysis as a tool for analyzing inhomogeneous aging: A case study for LiFePO4|Graphite cylindrical cells}},
url = {https://doi.org/10.1016/j.jpowsour.2017.09.059},
volume = {368},
year = {2017}
}

@article{FATH2019100813,
author = {Fath, Johannes Philipp and Dragicevic, Daniel and Bittel, Laura and Nuhic, Adnan and Sieg, Johannes and Hahn, Severin and Alsheimer, Lennart and Spier, Bernd and Wetzel, Thomas},
doi = {https://doi.org/10.1016/j.est.2019.100813},
issn = {2352-152X},
journal = {Journal of Energy Storage},
pages = {100813},
title = {{Quantification of aging mechanisms and inhomogeneity in cycled lithium-ion cells by differential voltage analysis}},
url = {https://www.sciencedirect.com/science/article/pii/S2352152X19301811},
volume = {25},
year = {2019}
}

@article{LEWERENZ2018421,
author = {Lewerenz, Meinert and Sauer, Dirk Uwe},
doi = {https://doi.org/10.1016/j.est.2018.06.003},
issn = {2352-152X},
journal = {Journal of Energy Storage},
pages = {421--434},
title = {{Evaluation of cyclic aging tests of prismatic automotive LiNiMnCoO2-Graphite cells considering influence of homogeneity and anode overhang}},
url = {https://www.sciencedirect.com/science/article/pii/S2352152X18301579},
volume = {18},
year = {2018}
}

@article{Sieg2020,
author = {Sieg, Johannes and Storch, Mathias and Fath, Johannes and Nuhic, Adnan and Bandlow, Jochen and Spier, Bernd and Sauer, Dirk Uwe},
doi = {10.1016/j.est.2020.101582},
issn = {2352152X},
journal = {Journal of Energy Storage},
number = {May},
pages = {101582},
publisher = {Elsevier},
title = {{Local degradation and differential voltage analysis of aged lithium-ion pouch cells}},
url = {https://doi.org/10.1016/j.est.2020.101582},
volume = {30},
year = {2020}
}

@ARTICLE{Lin2020-tb,
  title    = "Robust estimation of battery system temperature distribution
              under sparse sensing and uncertainty",
  author   = "Lin, Xinfan and Perez, Hector E and Siegel, Jason B and
              Stefanopoulou, Anna G",
  journal  = "IEEE Trans. Control Syst. Technol.",
  year     =  2020
}

@article{Marlow2024,
author = {{Naylor Marlow}, Max and Chen, Jingyi and Wu, Billy},
doi = {10.1038/s44172-023-00153-5},
issn = {2731-3395},
journal = {Communications Engineering},
number = {1},
pages = {2},
title = {{Degradation in parallel-connected lithium-ion battery packs under thermal gradients}},
url = {https://doi.org/10.1038/s44172-023-00153-5},
volume = {3},
year = {2024}
}

@article{Antti_Field,
author = {Aitio, Antti and Howey, David A},
doi = {https://doi.org/10.1016/j.joule.2021.11.006},
issn = {2542-4351},
journal = {Joule},
number = {12},
pages = {3204--3220},
title = {{Predicting battery end of life from solar off-grid system field data using machine learning}},
url = {https://www.sciencedirect.com/science/article/pii/S2542435121005328},
volume = {5},
year = {2021}
}

@article{DUBARRY201919,
title = {Battery energy storage system modeling: Investigation of intrinsic cell-to-cell variations},
journal = {Journal of Energy Storage},
volume = {23},
pages = {19-28},
year = {2019},
issn = {2352-152X},
doi = {https://doi.org/10.1016/j.est.2019.02.016},
url = {https://www.sciencedirect.com/science/article/pii/S2352152X18308156},
author = {Matthieu Dubarry and Carlos Pastor-Fernández and George Baure and Tung Fai Yu and W. Dhammika Widanage and James Marco}
}

@article{SULZER20211934,
author = {Sulzer, Valentin and Mohtat, Peyman and Aitio, Antti and Lee, Suhak and Yeh, Yen T and Steinbacher, Frank and Khan, Muhammad Umer and Lee, Jang Woo and Siegel, Jason B and Stefanopoulou, Anna G and Howey, David A},
doi = {https://doi.org/10.1016/j.joule.2021.06.005},
issn = {2542-4351},
journal = {Joule},
number = {8},
pages = {1934--1955},
title = {{The challenge and opportunity of battery lifetime prediction from field data}},
url = {https://www.sciencedirect.com/science/article/pii/S2542435121002932},
volume = {5},
year = {2021}
}

@inbook{Plett_textbook,
  author    = {Plett, Gregory},
  title     = {Battery Management Systems, Volume II: Equivalent-Circuit Methods},
  chapter   = {3},  
  publisher = {Artech},
  year      = {2015},
}

@article{WENG201336,
author = {Weng, Caihao and Cui, Yujia and Sun, Jing and Peng, Huei},
doi = {https://doi.org/10.1016/j.jpowsour.2013.02.012},
issn = {0378-7753},
journal = {Journal of Power Sources},
pages = {36--44},
title = {{On-board state of health monitoring of lithium-ion batteries using incremental capacity analysis with support vector regression}},
url = {https://www.sciencedirect.com/science/article/pii/S0378775313002668},
volume = {235},
year = {2013}
}

@article{Mohtat_2020,
doi = {10.1149/1945-7111/aba5d1},
url = {https://dx.doi.org/10.1149/1945-7111/aba5d1},
year = {2020},
month = {jul},
publisher = {IOP Publishing},
volume = {167},
number = {11},
pages = {110561},
author = {Mohtat, Peyman and Lee, Suhak and Sulzer, Valentin and Siegel, Jason B. and Stefanopoulou, Anna G.},
title = {Differential Expansion and Voltage Model for Li-ion Batteries at Practical Charging Rates},
journal = {Journal of The Electrochemical Society}
}

@article{Offer2012,
author = {Offer, Gregory J. and Yufit, Vladimir and Howey, David A. and Wu, Billy and Brandon, Nigel P.},
doi = {10.1016/j.jpowsour.2012.01.087},
issn = {03787753},
journal = {Journal of Power Sources},
pages = {383--392},
publisher = {Elsevier B.V.},
title = {{Module design and fault diagnosis in electric vehicle batteries}},
url = {http://dx.doi.org/10.1016/j.jpowsour.2012.01.087},
volume = {206},
year = {2012}
}

@article{HE2021102867,
author = {He, Zhigang and Shen, Xiaoyu and Sun, Yanyan and Zhao, Shichao and Fan, Bin and Pan, Chaofeng},
doi = {https://doi.org/10.1016/j.est.2021.102867},
issn = {2352-152X},
journal = {Journal of Energy Storage},
pages = {102867},
title = {{State-of-health estimation based on real data of electric vehicles concerning user behavior}},
url = {https://www.sciencedirect.com/science/article/pii/S2352152X21005892},
volume = {41},
year = {2021}
}

@ARTICLE{Weng2023-lj,
  title    = "Differential voltage analysis for battery manufacturing process
              control",
  author   = "Weng, Andrew and Siegel, Jason B and Stefanopoulou, Anna",
  journal  = "Frontiers in Energy Research",
  volume   =  11,
  year     =  2023
}

@article{Yuan2025,
author = {Yuan, Yuebo and Wang, Hewu and Sun, Yukun and Han, Xuebing and Zhu, Cheng and Ouyang, Minggao},
doi = {10.1016/j.energy.2025.135529},
issn = {18736785},
journal = {Energy},
number = {August 2024},
pages = {135529},
publisher = {Elsevier Ltd},
title = {{The influence of local lithium plating on battery safety and a novel detection method}},
url = {https://doi.org/10.1016/j.energy.2025.135529},
volume = {321},
year = {2025}
}

@article{Hu2020,
author = {Hu, Xiaosong and Zhang, Kai and Liu, Kailong and Lin, Xianke and Dey, Satadru and Onori, Simona},
doi = {10.1109/MIE.2020.2964814},
issn = {19410115},
journal = {IEEE Industrial Electronics Magazine},
number = {3},
pages = {65--91},
title = {{Advanced Fault Diagnosis for Lithium-Ion Battery Systems: A Review of Fault Mechanisms, Fault Features, and Diagnosis Procedures}},
volume = {14},
year = {2020}
}

@article{QI2024605,
author = {Qi, Qingguang and Liu, Wenxue and Deng, Zhongwei and Li, Jinwen and Song, Ziyou and Hu, Xiaosong},
doi = {https://doi.org/10.1016/j.jechem.2024.01.047},
issn = {2095-4956},
journal = {Journal of Energy Chemistry},
pages = {605--618},
title = {{Battery pack capacity estimation for electric vehicles based on enhanced machine learning and field data}},
url = {https://www.sciencedirect.com/science/article/pii/S2095495624000858},
volume = {92},
year = {2024}
}

@article{Figgener2024,
author = {Figgener, Jan and van Ouwerkerk, Jonas and Haberschusz, David and Bors, Jakob and Woerner, Philipp and Mennekes, Marc and Hildenbrand, Felix and Hecht, Christopher and Kairies, Kai Philipp and Wessels, Oliver and Sauer, Dirk Uwe},
doi = {10.1038/s41560-024-01620-9},
issn = {20587546},
journal = {Nature Energy},
number = {11},
pages = {1438--1447},
publisher = {Springer US},
title = {{Multi-year field measurements of home storage systems and their use in capacity estimation}},
url = {http://dx.doi.org/10.1038/s41560-024-01620-9},
volume = {9},
year = {2024}
}

@article{Feng2018a,
author = {Feng, Xuning and Pan, Yue and He, Xiangming and Wang, Li and Ouyang, Minggao},
doi = {10.1016/j.est.2018.04.020},
issn = {2352152X},
journal = {Journal of Energy Storage},
number = {April},
pages = {26--39},
publisher = {Elsevier},
title = {{Detecting the internal short circuit in large-format lithium-ion battery using model-based fault-diagnosis algorithm}},
url = {https://doi.org/10.1016/j.est.2018.04.020},
volume = {18},
year = {2018}
}

@article{Gao2019,
author = {Gao, Wenkai and Zheng, Yuejiu and Ouyang, Minggao and Li, Jianqiu and Lai, Xin and Hu, Xiaosong},
doi = {10.1109/TIE.2018.2838109},
issn = {02780046},
journal = {IEEE Transactions on Industrial Electronics},
number = {3},
pages = {2132--2142},
publisher = {IEEE},
title = {{Micro-short-circuit diagnosis for series-connected lithium-ion battery packs using mean-difference model}},
volume = {66},
year = {2019}
}

@article{en14144074,
author = {Movassagh, Kiarash and Raihan, Arif and Balasingam, Balakumar and Pattipati, Krishna},
doi = {10.3390/en14144074},
issn = {1996-1073},
journal = {Energies},
number = {14},
title = {{A Critical Look at Coulomb Counting Approach for State of Charge Estimation in Batteries}},
url = {https://www.mdpi.com/1996-1073/14/14/4074},
volume = {14},
year = {2021}
}

@article{XU2025115524,
author = {Xu, JiYang and Ma, Jian and Zhang, Kai and Li, JiaBo and Zhao, Xuan and He, ZongKe and Wu, XueQin},
doi = {https://doi.org/10.1016/j.est.2025.115524},
issn = {2352-152X},
journal = {Journal of Energy Storage},
pages = {115524},
title = {{SOC estimation and internal short circuit fault diagnosis based on DAEKF method for power batteries}},
url = {https://www.sciencedirect.com/science/article/pii/S2352152X25002373},
volume = {112},
year = {2025}
}

@article{Mohtat2022,
author = {Mohtat, Peyman and Lee, Suhak and Siegel, Jason B. and Stefanopoulou, Anna G.},
doi = {10.1016/j.jpowsour.2021.230714},
issn = {03787753},
journal = {Journal of Power Sources},
number = {August 2021},
publisher = {Elsevier B.V.},
title = {{Comparison of expansion and voltage differential indicators for battery capacity fade}},
volume = {518},
year = {2022}
}

@article{Song2020,
author = {Song, Ziyou and Delgado, Fanny Pinto and Hou, Jun and Hofmann, Heath and Sun, Jing},
doi = {10.23919/ACC45564.2020.9147423},
isbn = {9781538682661},
issn = {07431619},
journal = {Proceedings of the American Control Conference},
pages = {1155--1160},
title = {{Individual Cell Fault Detection for Parallel-Connected Battery Cells Based on the Statistical Model and Analysis}},
volume = {2020-July},
year = {2020}
}

@article{Zhao2024,
author = {Zhao, Jingyuan and Feng, Xuning and Tran, Manh Kien and Fowler, Michael and Ouyang, Minggao and Burke, Andrew F.},
doi = {10.1016/j.jpowsour.2024.234111},
issn = {03787753},
journal = {Journal of Power Sources},
number = {January},
title = {{Battery safety: Fault diagnosis from laboratory to real world}},
volume = {598},
year = {2024}
}

@article{Lin_resistance_operating_conditions,
author = {Lin, Xianke and Khosravinia, Kavian and Hu, Xiaosong and Li, Ju and Lu, Wei},
doi = {https://doi.org/10.1016/j.pecs.2021.100953},
issn = {0360-1285},
journal = {Progress in Energy and Combustion Science},
pages = {100953},
title = {{Lithium Plating Mechanism, Detection, and Mitigation in Lithium-Ion Batteries}},
url = {https://www.sciencedirect.com/science/article/pii/S0360128521000514},
volume = {87},
year = {2021}
}

@article{WENG_formation_prediction,
author = {Weng, Andrew and Mohtat, Peyman and Attia, Peter M and Sulzer, Valentin and Lee, Suhak and Less, Greg and Stefanopoulou, Anna},
doi = {https://doi.org/10.1016/j.joule.2021.09.015},
issn = {2542-4351},
journal = {Joule},
number = {11},
pages = {2971--2992},
title = {{Predicting the impact of formation protocols on battery lifetime immediately after manufacturing}},
url = {https://www.sciencedirect.com/science/article/pii/S2542435121004438},
volume = {5},
year = {2021}
}

@article{Pozzato2023,
author = {Pozzato, Gabriele and Allam, Anirudh and Pulvirenti, Luca and Negoita, Gianina Alina and Paxton, William A. and Onori, Simona},
doi = {10.1016/j.joule.2023.07.018},
issn = {25424351},
journal = {Joule},
number = {9},
pages = {2035--2053},
publisher = {Elsevier Inc.},
title = {{Analysis and key findings from real-world electric vehicle field data}},
url = {https://doi.org/10.1016/j.joule.2023.07.018},
volume = {7},
year = {2023}
}

@article{Cai2021,
author = {Cai, Ting and Valecha, Puneet and Tran, Vivian and Engle, Brian and Stefanopoulou, Anna and Siegel, Jason},
doi = {10.1016/j.etran.2020.100100},
issn = {25901168},
journal = {eTransportation},
title = {{Detection of Li-ion battery failure and venting with Carbon Dioxide sensors}},
year = {2021}
}

\end{document}